\documentclass[11pt]{article}    
\usepackage{amsmath,amssymb,amsfonts,latexsym,xspace,url}
\usepackage{color}
\newcommand{\relie}{\textsc{ReLie}\xspace}
\newcommand{\reduce}{\textsc{Reduce}\xspace}
\makeatletter
\g@addto@macro\@verbatim\small
\makeatother

\newtheorem{example}{Example}
\newtheorem{remark}{Remark}

\begin{document}

\title{\relie: a \reduce program
for Lie group analysis of differential equations}

\author{Francesco Oliveri\\
\ \\
{\footnotesize Department MIFT, University of Messina}\\
{\footnotesize Viale F. Stagno d'Alcontres 31, 98166 Messina, Italy}\\
{\footnotesize francesco.oliveri@unime.it; http://mat521.unime.it/oliveri}}

\date{}

\maketitle
\begin{abstract}
Lie symmetry analysis  provides a general theoretical framework for investigating ordinary and partial differential equations. The theory is completely algorithmic even if it usually involves lengthy computations. For this reason, many computer algebra packages have been developed along the years to automate the computation.
In this paper, we describe the program \relie, written in the Computer Algebra System \reduce, which since 2008 is an open source program (\url{http://www.reduce-algebra.com}) and is available for all platforms. 
\relie is able to perform almost automatically the needed computations for Lie symmetry analysis of
differential equations. Its source code is freely available at the url \url{http://mat521.unime.it/oliveri}. The use of the program is illustrated by means of some simple examples; nevertheless, it is to be underlined that it provides effective also for more complex computations where one has to deal with very large expressions.
\end{abstract}



\section{Introduction}
\label{sec:intro}
A general unified approach to deal with ordinary as well as partial differential equations is 
provided by the study of their continuous symmetries, \emph{i.e.}, transformations of their solution 
manifold into itself. Symmetry analysis of differential 
equations originated in the nineteenth century with Sophus Lie \cite{Lie1888book,Lie1891book} 
who recognized that many special integration theories of differential equations are a consequence 
of the invariance under one--parameter continuous groups of transformations. 
Such transformations establish a diffeomorphism on the space of independent and dependent 
variables, mapping solutions of the equations to other solutions. Any transformation of the 
independent and dependent variables in turn induces a transformation of the derivatives. 
As Lie himself showed, the problem of finding the Lie group of point transformations leaving a differential equation 
invariant consists in solving a \emph{linear} system of determining equations for the components of its infinitesimal 
generators. Lie's theory (many textbooks and monographies are available, for instance \cite{Ovsiannikov1982book,Ibragimov1985book,Olver1986book,%
BlumanKumei1989book,Stephani1989book,Ibragimov1994CRC1,%
Ibragimov1995CRC2,Ibragimov1996CRC3,%
Olver1995book,Baumann2000book,Cantwell2002book,BlumanAnco2002book,%
Meleshko2005book,BlumanCheviakovAnco2009book,Bordagbook})
allows for  the development of systematic procedures leading to the integration by quadrature 
(or at least to lowering the order) of ordinary differential equations \cite{Bluman:ode}, to the determination of invariant solutions of initial 
and boundary value problems 
\cite{Olver1986book,RogersAmes1989Book,Ibragimov1994CRC1,Ibragimov1995CRC2,%
OliveriSpeciale1998,OliveriSpeciale1999,OliveriSpeciale2002,%
Oliveri_ND2005,OliveriSpeciale2005}, to the derivation of conserved quantities 
\cite{Noether,Krasil-Vinogradov,Bluman-sigma}, to the algorithmic construction of invertible point transformations  mapping the differential equations into equivalent forms easier to handle
\cite{KumeiBluman1982,DonatoOliveri1994,DonatoOliveri1995,%
DonatoOliveri1996,CurroOliveri2008,Oliveri2010,Oliveri2012,%
GorgoneOliveri2017a,GorgoneOliveri2017b}, or to design efficient numerical schemes \cite{Olver:numerical,RebeloValiquette:numerical,%
Bihlo-Valiquette:numerical}, \ldots

Lie's classical theory is a source for various generalizations. Among these generalizations there are the nonclassical method first proposed by Bluman and Cole \cite{Bluman-Cole-1969}, and now part of the more general method of differential constraints \cite{Fuschich,Fuschich-Tsyfra}, the potential symmetries \cite{Bluman:potential}, the nonlocal symmetries \cite{Govinder-Leach:nonlocal,Leach-Andriopoulos:nonlocal}, the generalized symmetries \cite{Olver1986book}, which in turn generalize contact symmetries introduced by Lie himself, the equivalence transformations \cite{Ovsiannikov1982book,Lisle1992,IbragimovTorrisi1994,IbragimovTorrisiValenti,TorrisiTracinaproceedings1996,TorrisiTracina1998,OzerSuhubi}, to quote a few. A recent introduction is that of approximate symmetries 
\cite{BGI-1989,IbragimovKovalev,FS-1989,DSGO:lieapprox,Gorgone2018}  for differential equations containing small terms.

The application of Lie's theory to differential equations is completely algorithmic; nevertheless, it usually involves a lot of cumbersome and tedious calculations. Today, many powerful Computer Algebra Systems (CAS) (either commercial or open source) are available, and the needed algebraic manipulations can be rapidly done almost automatically. 
In fact, many specific packages for performing symmetry analysis of differential equations are currently available in the literature \cite{Schwarz1985,Schwarz1988,Head:muLie,Hereman1994,Sherring1993,%
Ibragimov1996CRC3,Hereman1997,Baumann2000book,CarminatiWu2000,Wolf,%
ButcherCasrminati2003,Cheviakov2007,Steeb2007book,Cheviakov2010,%
Sade2010,JeffersonCarminati2013}. 

In this paper, we describe the package \relie, written in the computer algebra system \reduce \cite{Reduce}. The routines contained in \relie allow the user to easily compute (exact and approximate) Lie point symmetries and conditional symmetries, as well as contact transformations, variational symmetries, and equivalence transformations for classes of differential equations containing arbitrary elements. 

\reduce is a general purpose computer algebra system whose development started by Anthony Hearn. It is written in a Lisp dialect (Portable Standard Lisp); since December 2008 it is an open source program (available  for all platforms at the url \url{http://www.reduce-algebra.com}).

In the following, we assume that every potential user of \relie is familiar with Lie symmetry analysis
of differential equations. Nevertheless, in order to fix the notation and keep the paper self--contained,  a brief sketch of the underlying theory is given.
The use of \relie requires the user to have a minimal experience in working with the algebraic mode of \reduce. Therefore, in this paper only the necessary details of the use of \relie will be presented. The source code of \relie, as well as the user's manual, can be found at the url 
\url{http://mat521.unime.it/oliveri}.

The origin of this package dates back to 1994 when the author developed some routines useful to manage the lengthy expressions needed to determine the Lie point symmetries of differential equations; this set of routines constantly grew through the years providing new capabilities, and now constitutes a package able to perform almost automatically much of the work. Moreover, the program can be used  also in interactive mode mimicking the steps one has to apply with pencil and paper but with the benefits of using a computer algebra system: this can represent  a useful  support in a higher course on symmetry analysis of differential equations or in all those situations (\emph{e.g.}, when one looks for conditional symmetries) where the determining equations can not be automatically solved and some special assumptions are needed.

The plan of the paper is the following. In Section~\ref{sec:theory}, we review the basic elements of Lie group theory of differential equations with reference to point symmetries (subsection~\ref{subsec:point}),  conditional symmetries (subsection~\ref{subsec:conditional}), contact symmetries (subsection~\ref{subsec:contact}),  variational symmetries (subsection~\ref{subsec:variational}),
approximate point, conditional, contact and variational symmetries (subsection~\ref{subsec:approximate}), and equivalence transformations (subsection~\ref{subsec:equivalence}). 
Then, in Section~\ref{sec:relie}, we consider some examples in order to illustrate the use of \relie. Finally, Section~\ref{sec:insiderelie} contains a description of the global variables and the main routines 
contained in the package.

\section{Basic elements of the theory}
\label{sec:theory}
\subsection{Lie point symmetries}
\label{subsec:point}
Within the framework of Lie group analysis, given a system of differential equations, say
\begin{equation}
\boldsymbol{\Delta}\left(\mathbf{x},\mathbf{u},\mathbf{u}^{(r)}\right)=\mathbf{0},
\label{sourcesystem}
\end{equation}
where $\mathbf{x}\in \mathcal{X}\subseteq\mathbb{R}^n$ is the set of the independent variables,
$\mathbf{u}\in \mathcal{U}\subseteq\mathbb{R}^m$ the set of the dependent variables,
and $\mathbf{u}^{(r)}$ the set of all (partial) derivatives of the $\mathbf{u}$'s with 
respect to the $\mathbf{x}$'s up to the order $r\ge 1$,
one is interested to find the admitted group of Lie point symmetries.
A Lie point symmetry is characterized by its infinitesimal generator 
\cite{Ovsiannikov1982book,Olver1986book,BlumanKumei1989book}
\begin{equation}
\Xi=\sum_{i=1}^n\xi_i(\mathbf{x}, \mathbf{u})\frac{\partial}{\partial x_i}+\sum_{\alpha=1}^m\eta_\alpha(\mathbf{x}, \mathbf{u})\frac{\partial}{\partial u_\alpha},
\label{vectorfield}
\end{equation}
where $\xi_i$ and $\eta_\alpha$ are the \emph{infinitesimals} of the group.
In dealing with differential equations, we need to extend (or prolong) the action of the Lie group to the $r$-th order \emph{jet space}, whose coordinates are the independent and dependent variables as well as the derivatives of the latter with respect to the former up to the order $r$.
Transformations for derivatives are obtained by requiring that the \emph{contact conditions} are preserved by the transformation (roughly speaking, the transformed derivatives have to be
the derivatives of the transformed dependent variables with respect to the transformed independent variables).

First order prolongation of the \emph{infinitesimal generator} reads
\[
\Xi^{(1)}=\Xi+\sum_{\alpha=1}^m\sum_{i=1}^n\eta_{[\alpha,i]}
\frac{\partial}{\partial u_{\alpha, i}},
\]
where
\[
\eta _{[\alpha,i]}=\frac{D\eta_\alpha}{Dx_i}-\sum_{j=1}^n u_{\alpha,j} \frac{D\xi_j}{Dx_i},
\]
and, in general, higher order extended infinitesimal generator has the form
\[
\Xi^{(k)}=\Xi^{(k-1)}+\sum_{\alpha=1}^m\sum_{i_1=1}^n\sum_{i_2=i_1}^n\ldots\sum_{i_k=i_{k-1}}^n\eta_{[\alpha,i_1\ldots i_k]}\frac{\partial}{\partial u_{\alpha,i_1\ldots i_k}},
\]
with
$\eta _{[\alpha,i_{1}\ldots  i_{k}]}$ defined recursively by the relation
\[
\eta _{[\alpha,i_{1}\ldots  i_{k}]}=\frac{D\eta_{[\alpha,i_{1}\ldots  i_{k-1}]}}{Dx_{i_k}}-\sum_{j=1}^n
u_{\alpha,ji_1 \ldots  i_{k-1}}\frac{D\xi_j}{Dx_{i_k}},
\]
the \emph{Lie derivative} being
\begin{equation}
\label{Liederivative}
\frac{D}{Dx_i}=\frac{\partial}{\partial x_i}+
\sum_{\alpha=1}^m \left(u_{\alpha,i}\frac{\partial}{\partial u_\alpha}+\sum_{j=1}^nu_{\alpha,ij} \frac{\partial}{\partial u_{\alpha,j}}+\dots\right),
\end{equation}
where $u_{\alpha,i}=\displaystyle\frac{\partial u_\alpha}{\partial x_i}$, 
$\displaystyle u_{\alpha,ij}=\frac{\partial^2 u_\alpha}{\partial x_i \partial x_j}$, $\ldots$

The point transformations leaving system~\eqref{sourcesystem} invariant are found by means of the straightforward Lie's algorithm,
which requires that the $r$-th order prolongation $\Xi^{(r)}$ 
of the vector field \eqref{vectorfield} acting on \eqref{sourcesystem} is zero along the solutions of
\eqref{sourcesystem}, \emph{i.e.},
\begin{equation}
\label{invariancecondition}
\left.\Xi^{(r)}\left(\boldsymbol\Delta\right)\right|_{\boldsymbol\Delta=\mathbf{0}}=\mathbf{0}.
\end{equation}

Notice that the  invariance condition \eqref{invariancecondition} involves polynomials in the variables
$\mathbf{u}^{(r)}$ if all these derivatives occur polynomially in the
differential equations: in fact, the prolonged infinitesimals are always 
polynomials in the derivatives $\mathbf{u}^{(r)}$.
Since we evaluate the invariance condition on the differential equations, the derivatives which appear in the invariance condition are independent. Therefore, it vanishes if and only if all the coefficients of such polynomials (involving the infinitesimals, their derivatives, the independent and the dependent variables) are zero, whereupon
we obtain an overdetermined set of linear and homogeneous partial differential equations (determining equations) for the infinitesimals $\boldsymbol\xi$ and 
$\boldsymbol\eta$, whose integration provides the generators of Lie point symmetries admitted by \eqref{sourcesystem}.
The infinitesimal generators of a Lie group, whose components are solutions of a linear homogeneous system of partial differential equations, span a vector space which can be finite or infinite--dimensional; by introducing an operation of commutation between two infinitesimal generators, this vector space gains the
structure of a Lie algebra \cite{deGraaf2000book,Erdmann2006book}.

\subsection{Q-conditional symmetries}
\label{subsec:conditional}
It is well known that some differential equations of interest in the applications possess very few Lie symmetries.
In 1969, Bluman and Cole \cite{Bluman-Cole-1969} introduced a generalization of classical Lie symmetries, and applied their method (called \emph{nonclassical}) to the linear heat equation. The basic idea was that of imposing the invariance to a system made by the differential equation at hand, the invariant surface condition and the differential consequences of the latter
\cite{Olver-Rosenau-1,Olver-Rosenau-2,Clarkson-Kruskal,Levi-Winternitz,NucciClarkson1992,Arrigo,Nucci1993,Clarkson-Mansfield,Saccomandi}.  
This method leads to \emph{nonlinear} determining equations whose general integration is usually difficult. Nonclassical symmetries are now part of conditional symmetries, \emph{i.e.}, symmetries of differential equations where some additional differential constraints are imposed to restrict the set of solutions;  some recent applications of conditional symmetries to reaction--diffusion equations can be found in \cite{Cherniha-JMAA2007,ChernihaPliukhin-JPA2008,Cherniha-JPA2010,ChernihaDavydovych-MCM2011,ChernihaDavydovych-CNSNS2012}.

Given a differential equation
\begin{equation}
\boldsymbol\Delta\left(\mathbf{x},\mathbf{u},\mathbf{u}^{(r)}\right)=0,
\end{equation}
and considering the invariant surface conditions $\mathbf{Q}=\mathbf{0}$, where
\begin{equation}
\mathbf{Q}= \sum_{i=1}^n \xi_i(\mathbf{x},\mathbf{u}) \frac{\partial\mathbf{u}}{\partial x_i}-
{\boldsymbol \eta}(\mathbf{x},\mathbf{u}),
\qquad \sum_{i=1}^n \xi_i^2 \neq 0,
\end{equation}
the Q-conditional symmetries are found by requiring
\begin{equation}
\left.\Xi^{(r)}(\boldsymbol\Delta)\right|_{\mathcal{M}}=0,
\end{equation}
where $\mathcal{M}$ is the manifold of the jet space defined by the system of equations
\begin{equation}
\boldsymbol\Delta=0, \qquad \mathbf{Q}=\mathbf{0}, \qquad \frac{\partial^{|k|} \mathbf{Q}}{\partial x_1^{k_1}\ldots\partial x_n^{k_n}}=\mathbf{0},
\end{equation}
where $1\le |k|=k_1+\ldots +k_n \le r-1$.

Of course,  a (classical) Lie symmetry is a (trivial) Q-conditional symmetry. However, differently from Lie symmetries, all possible Q-conditional symmetries of a differential equation form a set which is not a Lie algebra in the general case.
Moreover, if the infinitesimals of a Q-conditional symmetry are multiplied by an arbitrary smooth function we have still a Q-conditional symmetry. This means that we can look for Q-conditional symmetries in $n$ different situations:
\begin{itemize}
\item $\xi_1=1$; 
\item $\xi_i=1$ and $\xi_{j}=0$ for $1\le j <i \le n$.
\end{itemize}

When $m>1$ (more than one dependent variable), one can look for conditional symmetries of $q$-th type $(1\le q \le m)$ 
 \cite{Cherniha-JPA2010,ChernihaDavydovych-MCM2011,ChernihaDavydovych-CNSNS2012}, by requiring the condition
\begin{equation}
\left.\Xi^{(r)}(\boldsymbol\Delta)\right|_{\mathcal{M}_q}=\mathbf{0},
\end{equation}
where $\mathcal{M}_q$ is the manifold of the jet space defined by the system of equations
\begin{equation}
\begin{aligned}
&\boldsymbol\Delta=\mathbf{0}, \qquad Q_{i_1}=\ldots = Q_{i_q}=0, \\ 
&\frac{\partial^{|k|} {Q_{i_1}}}{\partial x_1^{k_1}\ldots\partial x_n^{k_n}}=\ldots = \frac{\partial^{|k|} {Q_{i_q}}}{\partial x_1^{k_1}\ldots\partial x_n^{k_n}}= 0,
\end{aligned}
\end{equation}
where $1\le |k|=k_1+\ldots +k_n \le r-1$, $1\le i_1<\cdots<i_q \le m$.
It is obvious that we may have in principle $\binom{m}{q}$ different sets of Q-conditional symmetries of $q$-th type, whereas for $q=0$ we simply have classical Lie symmetries. Moreover, when $q<m$ it may result easier to find solutions to the nonlinear determining equations.
Q-conditional symmetries allow for symmetry reductions of differential equations, and may provide explicit solutions not obtainable with classical symmetries.

\subsection{Contact transformations}
\label{subsec:contact}
Besides point transformations whose infinitesimals depend at most on independent and dependent variables, contact transformations  \cite{Eisenhart1928,Olver1986book,Lychagin2007book} play an important role in many applications. Nevertheless, true contact transformations exist for differential equations involving only one dependent variable (for more than one dependent variable, they are prolongations of point transformations), and their infinitesimal generator has the form
\begin{equation}
\label{symbol_contact}
\Xi = \sum_{i=1}^n \xi_i(\mathbf{x},u,\mathbf{u}^{(1)})\frac{\partial}{\partial x_i}+\eta(\mathbf{x},u,\mathbf{u}^{(1)})\frac{\partial}{\partial u}+\sum_{i=1}^n \eta_{[,i]}(\mathbf{x},u,\mathbf{u}^{(1)})\frac{\partial}{\partial u_{,i}}.
\end{equation}

The operator \eqref{symbol_contact} characterizes a group of contact transformations if and only if there exists a
function $\Omega(\mathbf{x},u,\mathbf{u}^{(1)})$ (characteristic function) such that
\begin{equation}
\xi_i=-\frac{\partial\Omega}{\partial u_{,i}}, \quad \eta = \Omega-\sum_{i=1}^n u_{,i}\frac{\partial\Omega}{\partial u_{,i}},\quad
\eta_{[,i]}=\frac{\partial\Omega}{\partial x_i}+u_{,i}\frac{\partial\Omega}{\partial u}.
\end{equation}

Prolongations of contact transformations for higher order derivatives are computed with the usual formulas.

Proper contact transformations (\emph{i.e.}, transformations that are not prolongations of point transformations) 
are the ones determined by a characteristic function $\Omega$ which is not linear in the first order derivatives.

\subsection{Variational symmetries}
\label{subsec:variational}
In dealing with differential equations, conservation laws have a deep relevance,
since they express the conservation of physical quantities such as mass, momentum, angular momentum,
energy, electrical charge. They are also important
due to their use in investigating integrability, existence, uniqueness and stability of solutions,
or in implementing efficient numerical methods  of integration
\cite{HairerLubichWanner2002,Iserlesetal2000}.

In 1918, Emmy Noether \cite{Noether} presented her celebrated procedure
(Noether's theorem) to find local conservation laws for differential equations arising
from  a variational principle. Noether proved that
a point symmetry of the action functional (action integral) provides
a local conservation law (\emph{i.e.}, a divergence expression) through an explicit formula
that involves the infinitesimals of the point symmetry and the Lagrangian itself.

Consider a functional given by an integral over a domain $\Omega\subset \mathbb{R}^n$, say
\begin{equation}
\mathcal{J}= \int_\Omega\mathcal{L}\left(\mathbf{x},\mathbf{u},\mathbf{u}^{(r)}\right)d\mathbf{x},
\end{equation}
where $\mathcal{L}(\mathbf{x},\mathbf{u},\mathbf{u}^{(r)})$ is a Lagrangian function.

By imposing the functional $\mathcal{J}$ to be stationary (for suitable variations of $\mathbf{u}$ and $\mathbf{u}^{(r)}$ vanishing on the boundary of the domain $\Omega$), we derive the Euler--Lagrange equations
\begin{equation}
E_{u_\alpha}(\mathcal{L})=0, \qquad \alpha=1,\ldots,m,
\end{equation}
where 
\begin{equation}
\label{euleroperator}
\begin{aligned}
E_{u_\alpha}(\cdot)&=\frac{\partial(\cdot)}{\partial u_\alpha}-\frac{D}{Dx_{j_1}}\left(\frac{\partial (\cdot)}{\partial
u_{\alpha,j_1}}\right)+\frac{D^2}{Dx_{j_1}Dx_{j_2}}\left(\frac{\partial (\cdot)}{\partial u_{\alpha,j_1j_2}}\right)\\
&-\ldots +(-1)^r\frac{D^r}{Dx_{j_1}\cdots Dx_{j_r}}\left(\frac{\partial(\cdot)}{\partial u_{\alpha,j_1\ldots j_r}}\right)
\end{aligned}
\end{equation}
is the Euler operator with respect to $u_\alpha$; here and in the following, if needed,  we use the Einstein convention 
of sums over repeated indices.

Noether's theorem \cite{Noether} states that if we know a Lie group of point transformations with generator $\Xi$
leaving the action integral invariant, and this holds true if the condition
\begin{equation}
\label{variational}
\Xi^{(r)}(\mathcal{L})+\mathcal{L}\sum_{i=1}^n\frac{D\xi_i}{Dx_i}=\sum_{i=1}^n\frac{D\phi_i}{Dx_i}
\end{equation}
for some functions $\phi_i(\mathbf{x},\mathbf{u},\mathbf{u}^{(r-1)})$ $(i=1,\ldots,n)$ is satisfied,
then the conservation law
\begin{equation}
\label{conservationlaw1}
\sum_{i=1}^n\frac{D}{Dx_i}\left(\xi_i\mathcal{L}+W_i-\phi_i\right)=0,
\end{equation}
where 
\begin{equation*}
\label{Wi}
\begin{aligned}
W_i&=
(\eta_\alpha-\xi_j u_{\alpha,j})\left(\frac{\partial \mathcal{L}}{\partial u_{\alpha,i}}+
\ldots + (-1)^{r-1}\frac{D^{r-1}}{D{x_{j_1}}\ldots D{x_{j_{r-1}}}}\left(\frac{\partial \mathcal{L}}{\partial
u_{\alpha,ij_1\ldots j_{r-1}}}\right)\right)\\
&+\frac{D(\eta_\alpha-\xi_j u_{\alpha,j})}{Dx_{j_1}}\left(\frac{\partial \mathcal{L}}{\partial u_{\alpha,ij_1}}+
\ldots + (-1)^{r-2}\frac{D^{r-2}}{D{x_{j_2}}\ldots D{x_{j_{r-1}}}}\left(\frac{\partial \mathcal{L}}{\partial u_{\alpha,ij_1\ldots j_{r-1}}}\right)\right)\\
&+\ldots+\frac{D^{r-1}(\eta_\alpha-\xi_j u_{\alpha,j})}{D{x_{j_1}}\ldots D{x_{j_{r-1}}}}\frac{\partial \mathcal{L}}
{\partial u_{\alpha,ij_1\ldots j_{r-1}}},
\end{aligned}
\end{equation*}
holds for any solution $\mathbf{u}(\mathbf{x})$ of Euler--Lagrange equations. For $n=1$, the conservation law provides a first integral of ordinary differential equations.

Boyer \cite{Boyer1967} extended Noether's theorem in order to
construct conservation laws arising from invariance under
generalized symmetries \cite{Olver1986book}, \emph{i.e.}, symmetries
with infinitesimals depending on higher order derivatives (see
\cite{BlumanCheviakovAnco2009book}).

\subsection{Approximate  symmetries}
\label{subsec:approximate}
Let 
\begin{equation}
\boldsymbol\Delta\left(\mathbf{x},\mathbf{u},\mathbf{u}^{(r)};\varepsilon\right)=0
\end{equation}
be a differential equation of order $r$ involving a small parameter $\varepsilon$. 
It is well known that any small perturbation of an equation usually destroys some important symmetries, and this restricts the applicability of Lie group methods to differential equations arising in concrete applications. On the other hand, differential equations containing \emph{small terms} are commonly and successfully investigated by means of perturbative techniques \cite{Nayfeh}. 
Therefore, it is desirable to combine Lie group methods with perturbation analysis, \emph{i.e.}, to  establish an approximate symmetry theory. In the literature, there are two widely used approaches to approximate symmetries: one has been proposed in 1988 by Baikov, Gazizov and Ibragimov \cite{BGI-1989} (see also \cite{IbragimovKovalev})  and another one 
in 1989 by Fushchich and Shtelen \cite{FS-1989}. Both approaches have pros and cons \cite{DSGO:lieapprox}. To overcome these problems, recently, a consistent approach has been proposed \cite{DSGO:lieapprox}, that is consistent with perturbation theory, and allows to extend all the relevant features of Lie group analysis to an approximate context. In what follows this latter approach is described.

If one looks for classical Lie point symmetries, in general it is not guaranteed that the infinitesimal generators depend on the parameter $\varepsilon$. Nevertheless, the occurrence of terms involving $\varepsilon$ has dramatic effects since one loses some symmetries admitted by the unperturbed equation
\begin{equation}
\boldsymbol\Delta\left(\mathbf{x},\mathbf{u},\mathbf{u}^{(r)};0\right)=0.
\end{equation}

Looking for solutions in the form
\begin{equation}
\label{expansion_u}
\mathbf{u}(\mathbf{x};\varepsilon)=\sum_{k=0}^p\varepsilon^k \mathbf{u}_{(k)}(\mathbf{x})+O(\varepsilon^{p+1}),
\end{equation}
the differential equation writes as
\begin{equation}
\boldsymbol\Delta\approx \sum_{k=0}^p\varepsilon^k\widetilde{\boldsymbol\Delta}_{(k)}\left(\mathbf{x},\mathbf{u}_{(0)},\mathbf{u}^{(r)}_{(0)},
\ldots,\mathbf{u}_{(k)},\mathbf{u}^{(r)}_{(k)}\right)=0.
\end{equation}
Now, let us consider a Lie generator
\begin{equation}
\Xi=\sum_{i=1}^n\xi_i(\mathbf{x},\mathbf{u};\varepsilon)\frac{\partial}{\partial x_i}
+\sum_{\alpha=1}^m\eta_\alpha(\mathbf{x},\mathbf{u};\varepsilon)\frac{\partial}{\partial u_\alpha},
\end{equation}
and assume that the infinitesimals  depend on the small parameter $\varepsilon$.

By using the expansion~\eqref{expansion_u} of the dependent variables only, we have for the infinitesimals
\begin{equation}
\xi_i\approx\sum_{k=0}^p\varepsilon^k \widetilde{\xi}_{(k)i}, \qquad \eta_\alpha\approx\sum_{k=0}^p\varepsilon^k\widetilde{\eta}_{(k)\alpha},
\end{equation}
where
\begin{equation}
\begin{aligned}
&\widetilde{\xi}_{(0)i}=\xi_{(0)i}=\xi_i(\mathbf{x},\mathbf{u}_{(0)},0),\qquad
&&\widetilde{\eta}_{(0)\alpha}=\eta_{(0)\alpha}=\eta_\alpha(\mathbf{x},\mathbf{u}_{(0)},0)\\
&\widetilde{\xi}_{(k+1)i}=\frac{1}{k+1}\mathcal{R}[\widetilde{\xi}_{(k)i}],\qquad &&\widetilde{\eta}_{(k+1)\alpha}=\frac{1}{k+1}\mathcal{R}[\widetilde{\eta}_{(k)\alpha}],
\end{aligned}
\end{equation}
$\mathcal{R}$ being a \emph{linear} recursion operator satisfying \emph{product rule} of derivatives and such that
\begin{equation}
\label{R_operator}
\begin{aligned}
&\mathcal{R}\left[\frac{\partial^{|\tau|}{f}_{(k)}(\mathbf{x},\mathbf{u}_{(0)})}{\partial u_{(0)1}^{\tau_1}\dots\partial u_{(0)m}^{\tau_m}}\right]=\frac{\partial^{|\tau|}{f}_{(k+1)}(\mathbf{x},\mathbf{u}_{(0)})}{\partial u_{(0)1}^{\tau_1}\dots\partial u_{(0)m}^{\tau_m}}\\
&\phantom{\mathcal{R}\left[\frac{\partial^{|\tau|}{f}_{(k)}(\mathbf{x},\mathbf{u}_{(0)})}{\partial u_{(0)1}^{\tau_1}\dots\partial u_{(0)m}^{\tau_m}}\right]}
+\sum_{i=1}^m\frac{\partial}{\partial u_{(0)i}}\left(\frac{\partial^{|\tau|} {f}_{(k)}(\mathbf{x},\mathbf{u}_{(0)})}{\partial u_{(0)1}^{\tau_1}\dots\partial u_{(0)m}^{\tau_m}}\right)u_{(1)i},\\
&\mathcal{R}[u_{(k)j}]=(k+1)u_{(k+1)j},
\end{aligned}
\end{equation}
where $k\ge 0$,  $j=1,\ldots,m$, $|\tau|=\tau_1+\cdots+\tau_m$.

Thence, we have an approximate Lie generator
\begin{equation}
\Xi\approx \sum_{k=0}^p\varepsilon^k\widetilde{\Xi}_{(k)},
\end{equation}
where
\begin{equation}
\widetilde{\Xi}_{(k)}=\sum_{i=1}^n\widetilde{\xi}_{(k)i}(\mathbf{x},\mathbf{u}_{(0)},\ldots,\mathbf{u}_{(k)})
\frac{\partial}{\partial x_i}
+\sum_{\alpha=1}^m\widetilde{\eta}_{(k)\alpha}(\mathbf{x},\mathbf{u}_{(0)},\ldots,\mathbf{u}_{(k)})\frac{\partial}{\partial u_\alpha}.
\end{equation}

Since we have to deal with differential equations, we need to prolong the Lie generator
to account for the transformation of derivatives. This is done as in classical Lie group analysis of differential equations,
\emph{i.e.}, the derivatives are transformed in such a way the contact conditions are preserved. 
Of course, in the expression of prolongations, we need to take into account the expansions of $\xi_i$, $\eta_\alpha$ and $u_\alpha$,  and drop the $O(\varepsilon^{p+1})$ terms.

\begin{example}
Let $p=1$, and consider the approximate Lie generator
\begin{equation}
\begin{aligned}
\Xi &\approx \sum_{i=1}^n\left(\xi_{(0)i}+\varepsilon\left(
 \xi_{(1)i}+\sum_{\beta=1}^m\frac{\partial \xi_{(0)i}}{\partial u_{(0)\beta}}u_{(1)\beta}\right)\right)\frac{\partial}{\partial x_i}\\
&+\sum_{\alpha=1}^m\left(\eta_{(0)\alpha}+\varepsilon\left(
 \eta_{(1)\alpha}+\sum_{\beta=1}^m\frac{\partial \eta_{(0)\alpha}}{\partial u_{(0)\beta}}u_{(1)\beta}\right)\right)\frac{\partial}{\partial u_\alpha},
 \end{aligned}
\end{equation}
where $\xi_{(0)i}$, $\xi_{(1)i}$, $\eta_{(0)\alpha}$ and $\eta_{(1)\alpha}$ depend on $(\mathbf{x},\mathbf{u}_{(0)})$.
The first order prolongation is
\begin{equation}
\Xi^{(1)}\approx\Xi + \sum_{\alpha=1}^m\sum_{i=1}^n \eta_{[\alpha,i]}\frac{\partial}{\partial u_{\alpha,i}},
\end{equation}
where
\begin{equation}
\begin{aligned}
\eta_{[\alpha,i]} &= \frac{\widehat{D}}{\widehat{D} x_i}\left(\eta_{(0)\alpha}+\varepsilon\left(
 \eta_{(1)\alpha}+\sum_{\beta=1}^m\frac{\partial \eta_{(0)\alpha}}{\partial u_{(0)\beta}}u_{(1)\beta}\right)\right)\\
 &-\sum_{j=1}^n \frac{\widehat{D}}{\widehat{D} x_i}\left(\xi_{(0)j}+\varepsilon\left(
 \xi_{(1)j}+\sum_{\beta=1}^m\frac{\partial \xi_{(0)j}}{\partial u_{(0)\beta}}u_{(1)\beta}\right)\right)
 \left(u_{(0)\alpha,j}+\varepsilon u_{(1)\alpha,j}\right),
\end{aligned}
\end{equation}
along with the Lie derivative
\begin{equation}
\frac{\widehat{D}}{\widehat{D} x_i}=\frac{\partial}{\partial x_i}+\sum_{k=0}^p\sum_{\alpha=1}^m \left(u_{(k)\alpha,i}
\frac{\partial}{\partial u_{(k)\alpha}}+\sum_{j=1}^n u_{(k)\alpha,ij} \frac{\partial}{\partial u_{(k)\alpha,j}}+\dots\right).
\end{equation}
Similar reasonings lead to higher order prolongations.
\end{example}

The approximate (at the order $p$) invariance condition of a differential equation reads
\begin{equation}
\left.\Xi^{(r)}\left(\boldsymbol\Delta\right)\right|_{\boldsymbol\Delta=O(\varepsilon^{p+1})}= O(\varepsilon^{p+1}).
\end{equation}
In the resulting condition we have to insert the expansion of $\mathbf{u}$ in order to obtain the determining 
equations at the various orders in $\varepsilon$.

The Lie generator $\widetilde{\Xi}_{(0)}$ is always a symmetry of the unperturbed equations ($\varepsilon=0$); the  \emph{correction}
terms $\displaystyle\sum_{k=1}^p\varepsilon^k\widetilde{\Xi}_{(k)}$ give
the deformation of the symmetry due to the terms involving $\varepsilon$. 
Not all symmetries of the unperturbed equations are admitted as the zeroth terms of approximate 
symmetries; the symmetries
of the unperturbed equations that are the zeroth terms of approximate symmetries are called \emph{stable 
symmetries} \cite{BGI-1989}.  

\begin{remark}
If $\Xi$ is the generator of an approximate Lie point symmetry of a differential equation, $\varepsilon\Xi$ is a generator of an
approximate Lie point symmetry too.
\end{remark}

By the same arguments as in classical Lie theory of differential equations it can be proved that
the approximate Lie point symmetries of a differential equation are the elements of an approximate Lie 
algebra  \cite{BGI-1989,IbragimovKovalev,DSGO:lieapprox}.

Of course, if one considers differential equations containing small terms, it is possible to construct approximate conditional symmetries \cite{GorgoneOliveriEJDE2018,GO-ZAMP}, the only difference with respect to exact symmetries being the structure of Lie generator 
which follows  the approach described above. In analogy with the exact case, for $p$-th order approximate conditional symmetries we have $n$ different cases:
\begin{itemize}
\item $\widetilde{\xi}_{(0)1}=1$, $\widetilde{\xi}_{(1)1}=\cdots=\widetilde{\xi}_{(p)1}=0$; 
\item $\widetilde{\xi}_{(0)i}=1$,  $\widetilde{\xi}_{(1)i}=\cdots=\widetilde{\xi}_{(p)i}=0$, and $\widetilde{\xi}_{(0)j}=\widetilde{\xi}_{(1)j}=\cdots = \widetilde{\xi}_{(p)j}=0$ for $1\le j <i \le n$.
\end{itemize}

We may also consider approximate contact transformations for scalar differential equations containing small terms by assuming the characteristic function $\Omega$ to depend on the small parameter $\varepsilon$. Limiting to first order approximate contact symmetries, the expansion for $\Omega$ reads
\begin{equation}
\begin{aligned}
\Omega&=\Omega_0\left(\mathbf{x},u_{(0)},u_{(0),i}\right)\\
&+\varepsilon\left(\Omega_1\left(\mathbf{x},u_{(0)},u_{(0),i}\right)
+\frac{\partial\Omega_0}{\partial u_{(0)}}u_{(1)}
+\sum_{i=1}^n \frac{\partial\Omega_0}{\partial u_{(0),i}}u_{(1),i}\right).
\end{aligned}
\end{equation}

Finally, also approximate variational symmetries for Lagrangians containing small terms can be defined.

\subsection{Equivalence transformations}
\label{subsec:equivalence}
There are situations where the differential equations at hand contain unspecified functions (\emph{arbitrary elements}); therefore, it is convenient to 
consider a class $\mathcal{E}(\mathbf{p})$ of differential equations involving some arbitrary continuously differentiable functions $p_k(\mathbf{x},\mathbf{u})$ ($k=1,\ldots,\ell$).

Let us consider a class of differential equations
\begin{equation}
 \boldsymbol\Delta\left(\mathbf{x},\mathbf{u},\mathbf{u}^{(r)};\mathbf{p},\mathbf{p}^{(s)}\right)=0,
\end{equation}
whose elements are given once we fix the functions $p_k$, where $\mathbf{p}^{(s)}$, denotes the set of all derivatives up to the order $s$ of the $\mathbf{p}$'s with respect to their arguments; these arguments 
may be the independent variables, the dependent ones and the derivatives of the latter with respect to the former
up to a fixed order $q$: let us denote the arguments of the arbitrary elements $\mathbf{p}$ as $\mathbf{z}\equiv(z_1,\ldots,z_N)$.

To face this problem, it is convenient to consider equivalence transformations, \emph{i.e.},
transformations that preserve the differential structure of the equations in the
system but may change the form of the constitutive functions and/or parameters
\cite{Ovsiannikov1982book,Lisle1992,IbragimovTorrisi1994,IbragimovTorrisiValenti,TorrisiTracinaproceedings1996,TorrisiTracina1998,OzerSuhubi}.

A one--parameter Lie group of equivalence transformations  \cite{Ovsiannikov1982book}
of a family $\mathcal{E}(\mathbf{p})$ of PDEs is a one--parameter Lie group of transformations
given by
\begin{equation} 
\mathbf{X}= \mathbf{X}(\mathbf{x}, \mathbf{u},\mathbf{p}; a), \qquad
\mathbf{U}= \mathbf{U}(\mathbf{x}, \mathbf{u},\mathbf{p}; a), \qquad
\mathbf{P} =\mathbf{P}(\mathbf{x}, \mathbf{u},\mathbf{p}; a),
\end{equation}
$a$ being the parameter, which is locally a $C^\infty$ diffeomorphism and maps a class $\mathcal{E}(\mathbf{p})$ of differential equations into itself; thus, it may change the differential equations (the form of the arbitrary elements therein involved) but preserves the differential structure.

A common assumption consists in considering transformations of the independent and dependent variables independent of the arbitrary elements $\mathbf{p}$.

In an \emph{augmented space}  $\mathcal{A}\equiv\mathcal{X}\times\mathcal{U}\times\mathcal{P}\subseteq\mathbb{R}^n  \times \mathbb{R}^m \times  \mathbb{R}^\ell$ \cite{Ovsiannikov1982book,Lisle1992}, where the independent variables, the dependent variables and the arbitrary functions live, respectively, the generator of the equivalence transformation,
\begin{equation}
\label{equivalenceoperator}
\Xi  =\sum_{i=1}^n\xi _{i}(\mathbf{x}, \mathbf{u})\frac{\partial}{\partial x_{i}}
+ \sum_{\alpha=1}^m\eta _{\alpha}(\mathbf{x}, \mathbf{u})\frac{\partial}{\partial u_\alpha}
+\sum_{j=1}^{\ell}\mu_{j}(\mathbf{x}, \mathbf{u},\mathbf{p})\frac{\partial}{\partial p_j},
\end{equation}
involves also the infinitesimals $\mu_{j}(\mathbf{x}, \mathbf{u},\mathbf{p})$ accounting for the arbitrary functions  $p_j$. The search for 
continuous equivalence transformations can be exploited by using the Lie's infinitesimal criterion \cite{Ovsiannikov1982book}.

The first prolongation of $\Xi$ writes as
\begin{equation}
\Xi^{(1)}= \Xi+
\sum_{\alpha=1}^m\sum_{i=1}^n \eta_{[\alpha,_i]} \frac{\partial}{\partial u_{\alpha,i}}+\sum_{j=1}^\ell \sum_{\beta=1}^{N} \mu_{[j,\beta]}\frac{\partial}{\partial p_{j,\beta}},
\end{equation}
with
\begin{equation}
\begin{aligned}
\eta_{[\alpha,i]}= \frac{D\eta_\alpha}{Dx_i}-\sum_{j=1}^n u_{\alpha,j}\frac{D\xi_j}{Dx_i},\qquad
\mu_{[j,\beta]}= \frac{\widetilde{D}\mu_j}{\widetilde{D}z_\beta}-\sum_{\gamma=1}^{N} p_{j,\gamma}\frac{\widetilde{D}\zeta_\gamma}{\widetilde{D}z_\beta},\\
\end{aligned}
\end{equation}
where $ p_{j,\gamma}=\displaystyle\frac{\partial p_j}{\partial z_\gamma}$ and $\zeta_\gamma$ is the infinitesimal generator of $z_\gamma$,
along with the \emph{Lie derivatives} $\displaystyle \frac{D}{Dx_i}$ already defined in \eqref{Liederivative} and
\begin{equation}
\frac{\widetilde{D}}{\widetilde{D}z_\beta}=\frac{\partial}{\partial z_\beta}+ \sum_{k=1}^\ell p_{k,\beta} \frac{\partial}{\partial p_k}.
\end{equation}
Higher order prolongations are immediately obtained by recursion,
\[
\begin{aligned}
\Xi^{(k)}&=\Xi^{(k-1)}+\sum_{\alpha=1}^m\sum_{i_1=1}^n\sum_{i_2=i_1}^n\ldots\sum_{i_k=i_{k-1}}^n\eta_{[\alpha,i_1\ldots i_k]}\frac{\partial}{\partial u_{\alpha,i_1\ldots i_k}}\\
&+\sum_{j=1}^\ell\sum_{\beta_1=1}^N\sum_{\beta_2=\beta_1}^N\ldots\sum_{\beta_k=\beta_{k-1}}^N\mu_{[j,\beta_1\ldots \beta_k]}\frac{\partial}{\partial p_{j,\beta_1\ldots\beta_k}},
\end{aligned}
\]
where
\begin{equation}
\mu _{[j,\beta_{1}\ldots  \beta_{k}]}=\frac{\widetilde{D}\mu_{[j,\beta_{1}\ldots  \beta_{k-1}}]}{\widetilde{D} z_{\beta_k}}-\sum_{\gamma=1}^N
p_{j,\gamma\beta_1\ldots\beta_{k-1}}\frac{\widetilde{D}\zeta_\gamma}{\widetilde{D} z_{\beta_k}}.
\end{equation}
In the augmented space $\mathcal{A}$, the arbitrary functions determining the class of differential equations
are assumed as dependent variables, and we require the invariance of the class in this augmented space via Lie's infinitesimal criterion \cite{Ovsiannikov1982book}:
\[
\left.\Xi^{(max(r,s))}\left(\boldsymbol\Delta\right)\right|_{\boldsymbol\Delta=0}=\mathbf{0}.
\]
If  we project the symmetries on the  space $\mathcal{Z}\equiv\mathcal{X}\times\mathcal{U}$ of the independent and dependent variables
(this is always possible if the infinitesimals of independent and dependent variables are assumed to be independent of $\mathbf{p}$),
we obtain a transformation changing an element of the class of differential equations to another element in the same class
(same differential structure but in general different arbitrary elements) \cite{OliveriSpeciale2012AAM,OliveriSpeciale2013JMP,GorgoneOliveriSpeciale2014}.
Such projected transformations map solutions
of a system in the class to solutions of a transformed system in the same class.

If some arbitrary elements do not depend upon some variables, the differential equations at hand have to be supplemented with \emph{auxiliary conditions}. For instance, the auxiliary conditions for the equivalence transformations of the class of differential equations
\cite{IbragimovTorrisiValenti}
\begin{equation}
\frac{\partial^2 u}{\partial t^2}-f\left(x,\frac{\partial u}{\partial x}\right)\frac{\partial^2 u}{\partial x^2}-g\left(x,\frac{\partial u}{\partial x}\right)=0,
\end{equation}
where $f$ and $g$ are unspecified functions of their arguments, are:
\begin{equation}
\frac{\partial f}{\partial t}=\frac{\partial f}{\partial u}=\frac{\partial f}{\partial (\partial u/\partial t)}=\frac{\partial g}{\partial t}=\frac{\partial g}{\partial u}=\frac{\partial g}{\partial (\partial u/\partial t)}=0.
\end{equation}

More in general, taking the transformations of the independent and dependent variables as functions of the arbitrary elements 
$\mathbf{p}$ too \cite{Meleshko:generalequivalence}, in the expression of prolongation we have to replace the Lie derivative $\displaystyle\frac{D}{Dx_i}$ with
\begin{equation}
\frac{D}{Dx_i}=\frac{\partial}{\partial x_i}+ \sum_{\alpha=1}^m u_{\alpha,i} \frac{\partial}{\partial u_\alpha}+\sum_{k=1}^\ell \left(\frac{\partial p_k}{\partial x_i}+\sum_{\beta=n+1}^N \frac{\partial p_k}{\partial z_\beta}\frac{\partial z_\beta}{\partial x_i}\right) \frac{\partial}{\partial p_k}.
\end{equation}

\subsection{Lie remarkable equations}
In \cite{MOV2007}, within the framework of inverse Lie problem, strongly and weakly Lie remarkable differential equations have been defined; relevant examples of such equations have been studied in \cite{MOV2007b,MOSV2014,GO2019_Lierem}. Their analysis involves the study of the rank of the distribution of prolongations of a Lie algebra of generators.

\section{The program \relie}
\label{sec:relie}
The application of Lie's group theory (together its generalization) to differential equations usually involves a lot of lengthy and cumbersome computations, so that the implementation of the algorithms in Computer Algebra Systems (CAS) is highly desired \cite{Schwarz1985,Schwarz1988,Head:muLie,Hereman1994,Sherring1993,%
Ibragimov1996CRC3,Hereman1997,Baumann2000book,CarminatiWu2000,Wolf,%
ButcherCasrminati2003,Cheviakov2007,Steeb2007book,Cheviakov2010,%
Sade2010,JeffersonCarminati2013}.

In this Section, we illustrate the use of \reduce \cite{Reduce} package \relie by considering some simple examples; more complex applications can be considered as well.  
The package \relie is a collection of (algebraic) routines that allow the user to investigate ordinary and partial differential equations 
within the framework of Lie symmetry analysis. By using the program, it is possible to compute almost automatically: 
Lie point symmetries, conditional symmetries, contact symmetries, variational symmetries (all these symmetries may be either exact or approximate) of differential equations, and equivalence transformations of classes of differential equations containing arbitrary elements. Moreover, the program implements functions for computing Lie brackets, the commutator table of a list of Lie generators, and the distribution of an algebra of Lie symmetries (useful in the context of inverse Lie problem \cite{MOV2007,MOV2007b,MOSV2014,GO2019_Lierem}). Remarkably, the program can be used interactively in all the cases where the determining equations are not automatically solved (for instance, when one looks for conditional symmetries or in some group classification problems).
 
After entering \reduce, assuming that we are in the directory containing the source file {\tt relie.red}, the set of
routines implemented in the package \relie becomes available after the statement
\begin{verbatim}
in "relie.red" $
\end{verbatim}
The loading of \relie is successful if the following output is displayed
\begin{verbatim}
ReLie, version 3.0
A REduce program for Lie group analysis of differential equations
(c) Francesco Oliveri (foliveri@unime.it) - 2020
Last update August 9, 2020
http://mat521.unime.it/oliveri/
\end{verbatim}
Alternatively, one may include the package in the \reduce image by issuing in a \reduce
session the statements
\begin{verbatim}
faslout 'relie $
in "relie.red" $
faslend $
\end{verbatim}
If all works, \relie is loaded through the command
\begin{verbatim}
load_package relie $
\end{verbatim}

According to the task we are interested to, some input data have to be provided to the program.
The \emph{minimal} required set of data is given by
a positive integer value for the global variable {\tt jetorder} (the maximum order of prolongation of the Lie generator), the list {\tt xvar} of the independent variables,
and the list {\tt uvar} of the dependent variables.

\begin{example}\label{example1}
The assignments
\begin{verbatim}
jetorder:=2 $
xvar:={t,x} $
uvar:={u} $
\end{verbatim}
refer to a second order partial differential equation for the unknown $u(t,x)$.
\end{example}  

The number of independent as well as dependent variables is arbitrary; moreover, the identifiers 
of independent and dependent variables are arbitrary provided that
no conflict arises with reserved words of \reduce or identifiers used by \relie (see 
Section~\ref{sec:insiderelie} for a list of internal global variables used by \relie). 
The value of {\tt jetorder} is \textbf{not} bounded by \relie.

The first function one has to call is {\tt relieinit()}, that
defines and initializes all the needed objects required to perform Lie group analysis.
It checks the input and possibly displays some warnings. If only {\tt jetorder}, {\tt xvar} and {\tt uvar} have been assigned, then calling {\tt relieinit()} produces the output:
\begin{verbatim}
The list 'diffeqs' of differential equation(s) is missing!
The list 'leadders' of leading derivative(s) is missing!
Check 'diffeqs' and/or 'leadders'!
You may only call relieprol() or generatealgebra(k) (k=1,2,3). 
\end{verbatim}

In fact, we did not define the list {\tt diffeqs} of differential equations and the list {\tt leadders} of
leading derivatives. Nevertheless, we can compute the {\tt jetorder}-th prolongation of the vector field generating a Lie group of point transformation.
The infinitesimal generators of the independent and dependent variables are automatically defined by the program (therefore, the user is not requested to set them), and stored in the list {\tt allinfinitesimals}. The infinitesimals of the independent variables are denoted by {\tt xi\_} followed by the identifiers in {\tt xvar}; analogously, the infinitesimals of the dependent variables are denoted by {\tt eta\_} followed by the identifiers in {\tt uvar}. 
According to the Example~\ref{example1}, the list  {\tt allinfinitesimals} is a list of two elements, each element being the list
{\tt \{xi\_t, xi\_x, eta\_u\}} (this redundancy is used in order to have a unified coding for dealing also with other kinds of symmetries, see below). The dependencies of the elements in {\tt allinfinitesimals} upon the independent  and dependent variables (the elements in {\tt xvar} and {\tt uvar}, respectively) 
are automatically set by \relie.

By calling the \relie function {\tt relieprol()}, as a result we get the list
{\tt prolongation}. This list has two elements: the first one is a list containing the coordinates of 
the jet space, the second one the list of the corresponding infinitesimal generators. Thus, the 
second element of {\tt prolongation} contains the components of the prolonged vector field up to 
{\tt jetorder}-th order.

\relie internally represents the derivatives of the dependent variables upon the independent variables by appending to the identifiers of the dependent variable the
underscore {\tt \_} followed by the identifiers of the involved independent variables.
For mixed derivatives the order of the independent variables in the internal representation of derivatives reflects the order in the list {\tt xvar}. For instance, if {\tt xvar} is {\tt \{t,x\}} and {\tt uvar} is {\tt\{u\}}, the left--hand side of the equation (Benjamin--Bono--Mahoney equation)
\begin{equation}
\frac{\partial u}{\partial t}+\frac{\partial u}{\partial x}+u\frac{\partial u}{\partial x}-\frac{\partial^3 u}{\partial x^2\partial t}=0
\end{equation}
has to be represented as
\begin{verbatim}
u_t + u_x + u*u_x - u_txx .
\end{verbatim}

\subsection{Computing Lie point symmetries of differential equations}
To compute the Lie point symmetries admitted by differential equations, in addition to
{\tt jetorder}, {\tt xvar} and {\tt uvar}, we have to provide:
\begin{enumerate}
\item the list {\tt diffeqs} of the left--hand sides of the differential equations
to be studied which are assumed with zero right--hand sides;
\item the list {\tt leadders} of some derivatives appearing in the differential equations: when computing the invariance conditions, the elements in the list {\tt leadders} are removed by solving the differential equations with respect to them.
\end{enumerate}

Both lists, {\tt diffeqs}   and {\tt leadders}, need to have the same length, otherwise a warning is displayed; moreover, the differential equations must be solvable with respect to the leading derivatives.
The user should be careful in the choice
of {\tt leadders}, especially in the case of systems of differential equations, in order
to guarantee that the differential equations stored in the list {\tt diffeqs} can be solved with respect to the elements in 
{\tt leadders}.

\begin{example}\label{exampleblasius}
Once \relie has been loaded, the statements
\begin{verbatim}
jetorder:=3 $
xvar:={x} $
uvar:={u} $
diffeqs:={u_xxx+u*u_xx/2} $
leadders:={u_xxx} $
relieinit() $
relieinv() $
reliedet() $
reliesolve() $
\end{verbatim}
allow to compute the Lie point symmetries of Blasius equation \cite{Blasius}
\[
\frac{d^3u}{dx^3}+\frac{1}{2}u\frac{d^2u}{dx^2}=0.
\]

The call to the function {\tt relieinit()} initializes the needed objects, then  the call to the function {\tt relieinv()}
provides the invariance condition of the differential equation defined in {\tt diffeqs}; in other terms, the 
{\tt jetorder}-th prolongation of the vector field of the Lie group is applied to the
differential equation and the equation itself is used to eliminate $\displaystyle\frac{d^3u}{dx^3}$. As a result, 
the list {\tt invcond} is provided:
\begin{verbatim}
invcond ->
   {6*df(eta_u,u,x,2)*u_x + 2*df(eta_u,u,x)*u*u_x 
    + 6*df(eta_u,u,x)*u_xx + 2*df(eta_u,u,3)*u_x**3 
    + 6*df(eta_u,u,2,x)*u_x**2 + df(eta_u,u,2)*u*u_x**2 
    + 6*df(eta_u,u,2)*u_x*u_xx + 2*df(eta_u,x,3) + df(eta_u,x,2)*u 
    - 6*df(xi_x,u,x2)*u_x**2 - 2*df(xi_x,u,x)*u*u_x**2 
    - 18*df(xi_x,u,x)*u_x*u_xx - 2*df(xi_x,u,3)*u_x**4 
    - 6*df(xi_x,u,2,x)*u_x**3 - df(xi_x,u,2)*u*u_x**3 
    - 12*df(xi_x,u,2)*u_x**2*u_xx + df(xi_x,u)*u*u_x*u_xx 
    - 6*df(xi_x,u)*u_xx**2 - 2*df(xi_x,x,3)*u_x 
    - df(xi_x,x,2)*u*u_x - 6*df(xi_x,x,2)*u_xx 
    + df(xi_x,x)*u*u_xx + eta_u*u_xx)}
\end{verbatim}
Since we have a scalar ordinary differential equation, the list {\tt invcond} contains
just one element, which is a polynomial in the variables {\tt u\_x} and 
{\tt u\_xx}. 
Invoking the function  {\tt reliedet()}, one gets the list
{\tt deteqs} of the coefficients of such (list of)  polynomial(s); thus, we have the \emph{determining equations}, \emph{i.e.},  a system of linear partial differential equations for the infinitesimals {\tt xi\_x} and {\tt eta\_u}:
\begin{verbatim}
deteqs ->
   {2*df(eta_u,x,3) + df(eta_u,x,2)*u,
    6*df(eta_u,u,x) - 6*df(xi_x,x,2) + df(xi_x,x)*u + eta_u,
    - 6*df(xi_x,u),
    6*df(eta_u,u,x,2) + 2*df(eta_u,u,x)*u - 2*df(xi_x,x,3) 
    - df(xi_x,x,2)*u,
    6*df(eta_u,u,2) - 18*df(xi_x,u,x) + df(xi_x,u)*u,
    6*df(eta_u,u,2,x) + df(eta_u,u,2)*u - 6*df(xi_x,u,x,2) 
    - 2*df(xi_x,u,x)*u,
    - 12*df(xi_x,u,2),
    2*df(eta_u,u,3) - 6*df(xi_x,u,2,x) - df(xi_x,u,2)*u,
    - 2*df(xi_x,u,3)}
\end{verbatim}

We may automatically solve this set of determining equations by calling the function
{\tt reliesolve()}, and the result turns out to be contained in the list {\tt symmetries}:
\begin{verbatim}
symmetries ->
{
 {
  {},
  {eta_u= - k_1*u,
   xi_x=k_1*x + k_2},
  {k_1,k_2},
  {}
 }
}
\end{verbatim}

The function {\tt reliesolve()} uses the \reduce package CRACK by S. Wolf \cite{Wolf}, 
specifically suitable to solve (overdetermined) systems of linear partial differential equations.

The list {\tt symmetries} contains an element (when more than one solution set is possible for particular choices of the involved parameters this list has one element for each solution set, see later)  which in turn is a list of 4 elements:
\begin{enumerate}
\item the first one is a list of conditions that are still unsolved (in this case the list is empty since no condition
remains unsolved);
\item the second one is a list giving the solution to the determining equations, \emph{i.e.}, the expressions of the infinitesimals;
\item the third one is a list containing the parameters involved in the solution 
(in this case {\tt k\_1} and {\tt k\_2});
\item the fourth one is a list of expressions which can not vanish (in this case the list is empty).
\end{enumerate}

The parameters involved in the symmetries of Blasius equation are constant; symmetries of partial differential equations may involve arbitrary functions that \relie denotes as {\tt f\_1}, {\tt f\_2}, \ldots; we can check the dependencies of such parameters by passing the name of the function to the procedure {\tt fargs()} (described in Crack \cite{Wolf}).

Since the infinitesimals are solutions of linear homogeneous differential equations, they are linear combinations of fundamental solutions. We can compute these fundamental solutions through the use of the function {\tt reliegen()}.
This function requires two arguments: an integer and a list. If {\tt reliesolve()} returns only one solution this integer is 1; when more than one solution is available (this may occur in group classification problems or in the search of conditional symmetries), the integer selects the solution. The second argument is a list of values. If the length of this list corresponds to the number of arbitrary (constant) parameters involved in the chosen symmetries, then a list of vector fields ({\tt generators}) is returned 
where the $i$-th parameter is replaced by the $i$-th element in the list and the other parameters are replaced by 0; 
otherwise,  the $i$-th parameter is replaced by 1 and the other parameters are replaced by 0.
For instance, invoking {\tt reliegen(1,\{1,1\})} (in this case we can write also {\tt reliegen(1,\{\})}), 
we get as a result the lists {\tt generators} and {\tt splitsymmetries}:
\begin{verbatim}
generators ->
{
 {x, -u},
 {1,0}
}
splitsymmetries ->
{
 {xi_x=x, eta_u=-u},
 {xi_x=1, eta_u=0}
}

\end{verbatim}
We have a list where each element is the list of the components of the vector field generating a symmetry (the components of the infinitesimals of the independent and dependent variables); the vector fields here characterize 
a scaling and a translation of the independent variable, respectively:
\[
\Xi_1=x\frac{\partial}{\partial x}-u\frac{\partial}{\partial u}, \qquad
\Xi_2=\frac{\partial}{\partial x}.
\]
\end{example}

\begin{remark}
By calling {\tt reliegen(1,\{-1\})}, the lists {\tt generators} and {\tt splitsymmetries} contain only one element representing the components of the infinitesimals in their general form, \emph{i.e.}, the linear combinations of all admitted generators.
For Blasius equation, the call {\tt reliegen(1,\{-1\})} provides
\begin{verbatim}
generators ->
{
 {k_1*x + k_2,- k_1*u}   
}
splitsymmetries ->
{
 {xi_x=k_1*x + k_2, eta_u=-k_1*u}
}
\end{verbatim}
\end{remark}

In the example above we detailed all the steps leading to the computation of the point symmetries of a differential equation. Nevertheless, we have a general function, called {\tt relie()}, requiring an integer argument: 
\begin{enumerate}
\item {\tt relie(1)} is equivalent to calling {\tt relieinit()};
\item {\tt relie(2)} is equivalent to calling in sequence {\tt relieinit()} and {\tt relieinv()};
\item {\tt relie(3)} is equivalent to calling in sequence {\tt relieinit()}, {\tt relieinv()} and 
{\tt reliedet()};
\item {\tt relie(4)} is equivalent to calling in sequence {\tt relieinit()}, {\tt relieinv()}, 
{\tt reliedet()} and {\tt reliesolve()}.
\end{enumerate}

\begin{example}\label{exampleheat}
After loading \relie, the assignments
\begin{verbatim}
jetorder:=2 $
xvar:={t,x} $
uvar:={u} $
diffeqs:={u_t-u_xx} $
leadders:={u_t} $
relie(4) $
\end{verbatim}
allow one  to easily compute the Lie point symmetries of the linear heat equation
\[
\frac{\partial u}{\partial t}-\frac{\partial^2 u}{\partial x^2}=0.
\]
The output produced by \relie is
\begin{verbatim}
symmetries ->
{
 {
  {df(f_1,t)-df(f_1,x,2)},
  {
   eta_u=(-8*f_1+2*k_1*u*x+2*k_4*t*u+k_4*u*x**2+2*k_6*u)/8,
   xi_x=(-k_1*t-k_2-k_3*x-k_4*t*x)/2,
   xi_t=(-2*k_3*t-k_4*t**2-2*k_5)/2
  },
  {f_1, k_1, k_2, k_3, k_4, k_5, k_6},
  {}
 }
}
\end{verbatim}

The infinitesimals of the point symmetries of the heat equation involve
7 parameters. The parameter {\tt f\_1} is a function, whereas the other parameters are constant.
By calling {\tt fargs(f\_1)}, we have as a result {\tt \{t,x\}}, denoting that the function {\tt f\_1} is a 
function of $t$ and $x$; moreover, this function satisfies the heat equation (the first element of the list {\tt symmetries}). The symmetry corresponding to this parameter
is responsible for the linear superposition principle of the solutions (typical of linear equations).

Calling the function {\tt reliegen(1,\{-2,-2,-2,-2,-1,4\})}, we get as a result the list {\tt generators} of vector fields of the finite subalgebra of the Lie algebra of symmetries of linear heat equation:
\begin{verbatim}
generators ->
{
 {0,t,( - u*x)/2},
 {0,1,0},
 {2*t,x,0},
 {t**2,t*x,(u*( - 2*t - x**2))/4}, 
 {1,0,0},
 {0,0,u} 
}
\end{verbatim}
We can reorder the vector fields in {\tt generators} by using the function {\tt newordering()}
\begin{verbatim}
generators:=newordering(generators,{5,2,6,3,1,4}) $
\end{verbatim}
so obtaining
\begin{verbatim}
generators ->
{
  {1,0,0},
  {0,1,0},
  {0,0, u}
  {2*t,x,0},
  {0, t,-u*x/2},
  {(t**2, t*x,u*(-2*t - x**2)/4},
}
\end{verbatim}
\emph{i.e.},
\[
\begin{aligned}
&\Xi_1=\frac{\partial}{\partial t},\qquad  \Xi_2=\frac{\partial}{\partial x}, \qquad \Xi_3=u\frac{\partial}{\partial u},\\
&\Xi_4=2t\frac{\partial}{\partial t}+x\frac{\partial}{\partial x},\qquad
\Xi_5=t\frac{\partial}{\partial x}-\frac{xu}{2}\frac{\partial}{\partial u},\\
&\Xi_6=t^2\frac{\partial}{\partial t}+xt\frac{\partial}{\partial x}
-\frac{u(2t + x^2)}{4}\frac{\partial}{\partial u},
\end{aligned}
\]
spanning a 6--dimensional Lie algebra.
The symmetry corresponding to the function {\tt f\_1},
\[
\Xi_f=f(t,x)\frac{\partial}{\partial u} \qquad \hbox{with}\qquad \frac{\partial f}{\partial t}-\frac{\partial^2 f}{\partial x^2}=0,
\]
renders the algebra of symmetries of linear heat equation infinite--dimensional.
\end{example}  

The function {\tt reliedet()} assumes that the invariance conditions are polynomials in the derivatives: this is true if all derivatives appear polynomially in the differential equations at hand.
In the case in which the differential equations we are investigating contain some derivatives not in polynomial form,
\relie constructs correctly the determining equations if we inform the program about the derivatives not occurring in polynomial form. These derivatives have to be the elements of the list {\tt nonpolyders}.

\begin{example}[Differential equations not polynomial in some derivatives]

The following code allows the user  to compute the Lie symmetries of the equation 
\[
\frac{\partial^2 u}{\partial t\partial x}+\exp\left(\frac{\partial u}{\partial x}\right)-u=0.
\]
\begin{verbatim}
jetorder:=2 $
xvar:={t,x} $
uvar:={u} $
diffeqs:={u_tx+exp(u_x)-u} $
leadders:={u_tx} $
nonpolyders:={u_x} $
relie(4) $
\end{verbatim}
\end{example}
\begin{remark}
The use of \relie is not limited to simple cases. For instance, consider the Navier-Stokes-Fourier equations of a gas in a rotating (with constant angular velocity $\omega$ along the vertical axis) frame reference and subject to gravity,
\begin{equation}
\begin{aligned}
&\frac{\partial\rho}{\partial t}+\sum_{k=1}^3\frac{\partial(\rho v_k)}{\partial x_k}=0,\nonumber\\
&\frac{\partial(\rho v_i)}{\partial t}+\sum_{k=1}^3\frac{\partial}{\partial
x_k} \left\{\rho v_iv_k-\frac{2}{\alpha}\left(\frac{K}{m}\right)T\left[\frac{1}{2}
\left(\frac{\partial v_i}{\partial x_k}+\frac{\partial
v_k}{\partial x_i}\right)
-\frac{1}{3}\frac{\partial v_r}{\partial x_r}\delta_{ik}\right]\right.\\
&\quad \left.+\rho\left(\frac{K}{m}\right)T\delta_{ik}\right\}=\rho F_i,\qquad i=1,2,3,\\
&\frac{\partial (\rho T)}{\partial t}+\frac{\partial}{\partial
x_k}\left\{
\rho T v_k-\frac{5}{2\alpha}\left(\frac{K}{m}\right)T\frac{\partial T}{\partial x_k}\right\}\nonumber\\
&\quad +\sum_{j,k=1}^3\left\{\frac{2}{3}\rho T\delta_{jk}-\frac{4}{3\alpha}T
\left[\frac{1}{2} \left(\frac{\partial v_j}{\partial
x_k}+\frac{\partial v_k}{\partial x_j}\right)
-\sum_{\ell=1}^3\frac{1}{3}\frac{\partial v_\ell}{\partial
x_\ell}\delta_{jk}\right]\right\}\frac{\partial v_j}{\partial x_k}=0,
\end{aligned}
\end{equation}
where $\rho$ is the mass density, $\mathbf{v}\equiv(v_1,v_2,v_3)$ the velocity, $T$ the temperature, $F_i$ the $i$-th component of the sum of external and inertial force,
\[
\mathbf{F}=(\rho(2\omega
v_2+\omega^2 x_1), \rho(-2\omega v_1+\omega^2 x_2),-\rho g),
\]
$g$ the gravitational acceleration, $K$ the Boltzmann
constant, $m$ the mass of a single particle, and the coefficient $\alpha$ is
an appropriate constant which describes the interaction between the Maxwellian molecules.
The computation of the Lie symmetries (spanning a 12th
dimensional Lie algebra) is done automatically in few second by means of \relie.
\end{remark}

\subsubsection{An example of group classification}
\relie is able to face the problem of group classification of differential equations. In fact, in certain cases it
may be interesting to check if special instances of arbitrary constants and/or functions involved in the differential equations lead to different sets of Lie symmetries.
This problem is faced by including the parameters we want to check in the list {\tt freepars}, as illustrated
in the following example. Moreover, we may define the list {\tt nonzeropars} containing parameters 
(constants or functions involved in the differential equations) or expressions 
that can not be vanishing.
\begin{example}\label{euler3d}
Consider the 3D Euler equations of ideal gas dynamics,
\[
\begin{aligned}
&\frac{\partial \rho}{\partial t}+\sum_{k=1}^3 v_k\frac{\partial\rho}{\partial x_k}+\rho\sum_{k=1}^3\frac{\partial v_k}{\partial x_k}=0,\\
&\rho\left(\frac{\partial v_i}{\partial t}+\sum_{k=1}^3 v_k\frac{\partial v_i}{\partial x_k}\right)+\frac{\partial p}{\partial x_i}=0,\qquad i=1,2,3,\\
&\frac{\partial p}{\partial t}+\sum_{k=1}^3 v_k\frac{\partial p}{\partial x_k}+\Gamma p\sum_{k=1}^3\frac{\partial v_k}{\partial x_k}=0,
\end{aligned}
\]
where $\rho(t,\mathbf{x})$ is the mass density, $v_i(t,\mathbf{x})$ the components of velocity, $p(t,\mathbf{x})$ the pressure, 
and $\Gamma$ the adiabatic exponent.
The following code performs the required task:
\begin{verbatim}
jetorder:=1 $
xvar:={t,x1,x2,x3} $
uvar:={rho,v1,v2,v3,p} $
freepars:={gamma} $
nonzeropars:={gamma} $
diffeqs:={rho_t+v1*rho_x1+v2*rho_x2+v3*rho_x3+rho*(v1_x1+v2_x2+v3_x3),
          rho*(v1_t+v1*v1_x1+v2*v1_x2+v3*v1_x3)+p_x1,
          rho*(v2_t+v1*v2_x1+v2*v2_x2+v3*v2_x3)+p_x2,
          rho*(v3_t+v1*v3_x1+v2*v3_x2+v3*v3_x3)+p_x3,
          p_t+v1*p_x1+v2*p_x2+v3*p_x3+gamma*p*(v1_x1+v2_x2+v3_x3)} $
leadders:={rho_t,v1_t,v2_t,v3_t,p_t} $
relie(4) $
\end{verbatim}
As a result, we get 
\begin{verbatim}
symmetries ->
{
 {
  {},
  {
   gamma=5/3,
   eta_p=5*k_10*p*t -k_14*p,
   eta_v3=k_10*t*v3-k_10*x3+k_12*v3+k_3*v2-k_5*v1-k_7-k_8*v3,
   eta_v2=-k_1+k_10*t*v2-k_10*x2+k_12*v2-k_2*v1-k_3*v3-k_8*v2,
   eta_v1=k_10*t*v1-k_10*x1-k_11+k_12*v1+k_2*v2+k_5*v3-k_8*v1,
   eta_rho=3*k_10*rho*t-2*k_12*rho-k_14*rho+2*k_8*rho,
   xi_x3=-k_10*t*x3+k_3*x2-k_5*x1-k_6-k_7*t-k_8*x3,
   xi_x2=-k_1*t-k_10*t*x2-k_2*x1-k_3*x3-k_4-k_8*x2,
   xi_x1=-k_10*t*x1-k_11*t+k_2*x2+k_5*x3-k_8*x1-k_9,
   xi_t=-k_10*t**2-k_12*t-k_13
  },
  {
   k_1, k_2, k_3, k_4, k_5, k_6, k_7, k_8, k_9, k_10, k_11, k_12, k_13, k_14
  },
  {}
 },
 {
  {},
  {
   eta_p=-k_14*p,
   eta_v3=k_12*v3+k_3*v2-k_5*v1-k_7-k_8*v3,
   eta_v2=-k_1+k_12*v2-k_2*v1-k_3*v3-k_8*v2,
   eta_v1=-k_11+k_12*v1+k_2*v2+k_5*v3-k_8*v1,
   eta_rho=-2*k_12*rho-k_14*rho+2*k_8*rho,
   xi_x3=k_3*x2-k_5*x1-k_6-k_7*t-k_8*x3,
   xi_x2=-k_1*t-k_2*x1-k_3*x3-k_4-k_8*x2,
   xi_x1=-k_11*t+k_2*x2+k_5*x3-k_8*x1-k_9,
   xi_t=-k_12*t-k_13
  },
  {
    gamma, k_1, k_2, k_3, k_4, k_5, k_6, k_7, k_8, k_9, k_11, k_12, k_13, k_14
  },
  {
    gamma, 3*gamma - 5
  }
 }
}
\end{verbatim}
In this case, {\tt symmetries} is a list of 2 elements since we have different solutions to the determining equations according to the value of {\tt gamma} (included in the list {\tt freepars}). 
Looking at the results provided by \relie, the first solution set refers to  $\Gamma=5/3$, the second one to a value of $\Gamma$ different from 0 and $5/3$.
By calling 
\begin{verbatim}
reliegen(2,{-1,1,-1,-1,1,-1,-1,-1,-1,-1,-1,-1,-1}) $
generators:=newordering(generators,{12,9,4,6,2,5,3,10,1,7,11,8,13}) $
\end{verbatim}
we obtain the following result
\begin{verbatim}
generators ->
{
 {1,0,0,0,0,0,0,0,0},
 {0,1,0,0,0,0,0,0,0},
 {0,0,1,0,0,0,0,0,0},
 {0,0,0,1,0,0,0,0,0},
 {0,x2, - x1,0,0,v2, - v1,0,0},
 {0,x3,0, - x1,0,v3,0, - v1,0},
 {0,0,x3, - x2,0,0,v3, - v2,0},
 {0,t,0,0,0,1,0,0,0},
 {0,0,t,0,0,0,1,0,0},
 {0,0,0,t,0,0,0,1,0},
 {t,0,0,0,2*rho,-v1,-v2,-v3,0},
 {0,x1,x2,x3,-2*rho,v1,v2,v3,0},
 {0,0,0,0,rho,0,0,0,p}
}
\end{verbatim}
so that  for $\Gamma$ unspecified we have  a 13--dimensional Lie algebra of point symmetries 
spanned by
\[
\begin{aligned}
&\Xi_1=\frac{\partial}{\partial t},  \qquad \Xi_{1+k}=\frac{\partial}{\partial x_k}, \qquad \Xi_{4+k}=t\frac{\partial}{\partial x_k}+\frac{\partial}{\partial v_k}, \qquad k=1,2,3, \\
&\Xi_{5+i+j}=x_j\frac{\partial}{\partial x_i}-x_i\frac{\partial}{\partial x_j}+v_j\frac{\partial}{\partial v_i}-v_i\frac{\partial}{\partial v_j}, \qquad 1\le i<j\le 3,\\
&\Xi_{11}=t\frac{\partial}{\partial t}+2\rho\frac{\partial}{\partial\rho}-\sum_{k=1}^3v_k\frac{\partial}{\partial v_k},\qquad
\Xi_{12}=\sum_{k=1}^3\left(x_k\frac{\partial}{\partial x_k}+v_k\frac{\partial}{\partial v_k}\right)-2\rho\frac{\partial}{\partial\rho},\\
&\Xi_{13}=\rho\frac{\partial}{\partial \rho}+p\frac{\partial}{\partial p}.
\end{aligned}
\] 
Calling
\begin{verbatim}
reliegen(1,{-1,1,-1,-1,1,-1,-1,-1,-1,-1,-1,-1,-1,-1}) $
generators:=newordering(generators,{13,9,4,6,2,5,3,11,1,7,12,8,14,10}) $
\end{verbatim}
we have that for $\Gamma=5/3$ there is a fourteenth symmetry generated by
\[
\Xi_{14}=t^2\frac{\partial}{\partial t}+\sum_{k=1}^3 \left(x_k t\frac{\partial}{\partial x_k}+(x_k-v_kt)\frac{\partial}{\partial v_k}\right)-3\rho t\frac{\partial}{\partial \rho}-5pt\frac{\partial}{\partial p}.
\]
\end{example}

\begin{remark}
The differential equations can contain arbitrary functions (in this case we need to 
declare in \reduce the variables these functions depend on), and they can be elements of the list {\tt freepars} 
and/or {\tt nonzeropars}.

For instance, if we want to classify the symmetries admitted by the Korteweg--deVries equation with variable coefficients \cite{Oliveri1987}
\[
\frac{\partial u}{\partial t}+u\frac{\partial u}{\partial x}+\frac{\partial^3 u}{\partial x^3}+f(t)u=0,
\]
where $f(t)$ is an unspecified function of time, the  input data for \relie are the following ones:
\begin{verbatim}
jetorder:=3 $
xvar:={t,x} $
uvar:={u} $
depend f,t $
freepars:={f} $
diffeqs:={u_t+u*u_x+u_xxx+f*u} $
leadders:={u_t} $
relie(4) $
\end{verbatim} 
\end{remark}

\subsubsection{Commutator table}
\relie is able to compute the commutator table of a set of infinitesimal generators spanning a Lie algebra.
Let us illustrate how to do with an example.
\begin{example}\label{burgers}
Consider the Burgers' equation
\[
\frac{\partial u}{\partial t}+u\frac{\partial u}{\partial x}-\frac{\partial^2 u}{\partial x^2}=0.
\]
To compute its Lie symmetries we provide to \relie the following input:
\begin{verbatim}
jetorder:=2 $
xvar:={t,x} $
uvar:={u} $
diffeqs:={u_t+u*u_x-u_xx} $
leadders:={u_t} $
relie(4) $
\end{verbatim}
We get quickly the output
\begin{verbatim}
symmetries ->
{
 {
  {},
  {
   {0, - k_1*t, - k_1},
   {0, - k_2,0},
   { - k_3*t,( - k_3*x)/2,(k_3*u)/2},
   {( - k_4*t**2)/2,( - k_4*t*x)/2,(k_4*(t*u - x))/2},
   { - k_5,0,0}
  },
  {}
 }
}
\end{verbatim}
Therefore, we have a 5--dimensional Lie algebra of point symmetries.
After extracting and reordering the components of the infinitesimal generators,
\begin{verbatim}
reliegen(1,{-1,-1,-2,-2,-1}) $
generators:=newordering(generators,{5,2,1,3,4}) $
\end{verbatim}
we have
\begin{verbatim}
generators ->
  {
   {1, 0, 0},
   {0, 1,0},
   {0, t, 1},
   {2*t, x, -u},
   {t**2, t*x, -t*u + x)}
  }
\end{verbatim}
that is a 5--dimensional Lie algebra spanned by
\begin{equation}
\label{symmetries-burgers}
\begin{aligned}
&\Xi_1=\frac{\partial}{\partial t}, \qquad \Xi_2=\frac{\partial}{\partial x},\qquad 
\Xi_3=t\frac{\partial}{\partial x}+\frac{\partial}{\partial u},\\
&\Xi_4=2t\frac{\partial}{\partial t}+x\frac{\partial}{\partial x}-u\frac{\partial}{\partial u}, \qquad
\Xi_5=t^2\frac{\partial}{\partial t}+xt\frac{\partial}{\partial x}+(x-tu)\frac{\partial}{\partial u}.
\end{aligned}
\end{equation}
The commutator table of the generators \eqref{symmetries-burgers} is obtained by calling
\begin{verbatim}
commutatortable(generators);
\end{verbatim}
that produces the output
\begin{verbatim}
commtable ->
{
 {0, 0, -vf_1, -vf_5, 0},
 { 0, 0, -2*vf_2, -vf_3, -vf_1},
 {vf_1, 2*vf_2, 0, -2*vf_4, -vf_5},
 {vf_5, vf_3, 2*vf_4, 0, 0},
 {0, vf_1, vf_5, 0, 0}
}
\end{verbatim}
where the five elements of the list {\tt generators} have been referred to as {\tt vf\_1}, {\tt vf\_2},
{\tt vf\_3}, {\tt vf\_4}, {\tt vf\_5}, respectively.
\end{example}

\subsection{Computation of conditional symmetries}
Conditional symmetries are computed by using the same functions as above; we only need to assign
a non--zero value (in the range between 1 and the number of independent variables) to the variable {\tt nonclassical}  and define the list {\tt qcond} (a list of distinct integers chosen in the range from 1 to the number of dependent variables).
Suppose that {\tt xvar:=\{x1,x2,x3\}} and {\tt uvar:=\{u,v\}}. 
The components of the Lie generator of the conditional symmetries are
\begin{itemize}
\item {\tt \{1, xi\_x2, xi\_x3, eta\_u, eta\_v\}} if {\tt nonclassical = 1};
\item {\tt \{0, 1, xi\_x3, eta\_u, eta\_v\}} if {\tt nonclassical = 2};
\item {\tt \{0, 0, 1, eta\_u, eta\_v\}} if {\tt nonclassical = 3}.
\end{itemize}

The list {\tt qcond} tells to the program which invariant surface conditions have to be used to restrict the solution manifold:
\begin{itemize}
\item {\tt qcond:=\{1,2\}} if the invariant surface conditions of both dependent variables have to be used;
\item {\tt qcond:=\{1\}} if the invariant surface condition of the first dependent variable has to be used;
\item {\tt qcond:=\{2\}} if the invariant surface condition of the second dependent variable has to be used.
\end{itemize}

Looking for conditional symmetries leads to \emph{nonlinear} determining equations, whereupon it is not unusual that {\tt reliesolve()} may fail to recover the solution. In such cases, it is necessary to make some special assumptions 
on the infinitesimals and/or try to solve interactively the determining equations.

\begin{example}{Nonclassical symmetries of linear heat equation}

The code
\begin{verbatim}
jetorder:=2 $
xvar:={t,x} $
uvar:={u} $
nonclassical:=1 $
qcond:={1} $
diffeqs:={u_t-u_xx} $
leadders:={u_xx} $
relie(4) $
\end{verbatim}
immediately computes the conditional symmetries, corresponding to the operator
\begin{equation}
\Xi=\frac{\partial}{\partial t}+\xi(t,x,u)\frac{\partial}{\partial x}+\eta(t,x,u)\frac{\partial}{\partial u},
\end{equation}
of the linear heat equation
\begin{equation}
\frac{\partial u}{\partial t}-\frac{\partial^2 u}{\partial x^2}=0.
\end{equation}
As a result, we get
\begin{verbatim}
symmetries ->
{
 {
  {df(f_1,t) - df(f_1,x,2) - 2*df(f_2,x,2)*f_1,
   2*df(f_2,t,x,2) + 2*df(f_2,t,x)*df(f_2,x) - df(f_2,t,2) 
     + 2*df(f_2,t)*df(f_2,x,2) - df(f_2,x,4) 
     - 2*df(f_2,x,3)*df(f_2,x) - 4*df(f_2,x,2)**2 
     - 2*df(f_2,x,2)*df(f_2,x)**2
  },
  {
   eta_u=(df(f_2,t)*u-df(f_2,x,2)*u-df(f_2,x)**2*u-2*f_1)/2,
   xi_x= - df(f_2,x)
  },
  {
   f_1,f_2
  },
  {}
 }
}
\end{verbatim}
Therefore, we have:
\begin{equation}
\xi = -\frac{\partial f_2}{\partial x}, \qquad \eta = -f_1 +\frac{1}{2}\left(\frac{\partial f_2}{\partial t}-\frac{\partial^2 f_2}{\partial x^2}-\left(\frac{\partial f_2}{\partial x}\right)^2\right)u,
\end{equation}
where $f_1(t,x)$ and $f_2(t,x)$ satisfy the differential equations
\begin{equation}
\begin{aligned}
&\frac{\partial f_1}{\partial t}-\frac{\partial^2 f_1}{\partial x^2}-2f_1\frac{\partial^2 f_2}{\partial x^2}=0,\\
&2\frac{\partial^3 f_2}{\partial t\partial x^2}+2\frac{\partial^2 f_2}{\partial t\partial x}\frac{\partial f_2}{\partial x}
-\frac{\partial^2 f_2}{\partial t^2}+2\frac{\partial f_2}{\partial t}\frac{\partial^2 f_2}{\partial x^2}-\frac{\partial^4 f_2}{\partial x^4}
-2\frac{\partial f_2}{\partial x}\frac{\partial^3 f_2}{\partial x^3}\\
&\qquad+4\left(\frac{\partial^2 f_2}{\partial x^2}\right)^2-2\left(\frac{\partial f_2}{\partial x}\right)^2\frac{\partial^2 f_2}{\partial x^2}=0.
\end{aligned}
\end{equation}
\end{example}

\begin{remark}
When computing conditional symmetries we have to be careful in the 
choice of {\tt leadders}, since 
also the invariant surface conditions and their differential consequences are 
solved with respect to some derivatives.
\end{remark}

\subsection{Computation of contact symmetries}
For computing contact symmetries of a scalar differential equation we need to assign the value 1
to the variable {\tt contact}. The functions that we have to call do not change, as the following example shows.
\begin{example}
Let us compute the contact symmetries of the ordinary differential equation
\[
\frac{d^3u}{dx^3}=0.
\]
The code is as follows.
\begin{verbatim}
jetorder:=3 $
xvar:={x} $
uvar:={u} $
contact:=1 $
diffeqs:={u_xxx} $
leadders:={u_xxx} $
relie(4) $
\end{verbatim}
As a result we get:
\begin{verbatim}
symmetries ->
{
 {
  {},
  {omega=(-2*k_1*x**2-4*k_10*u_x-4*k_2*x-4*k_3*u-4*k_4+4*k_5*u*u_x 
    -2*k_5*u_x**2*x-4*k_6*u**2+4*k_6*u*u_x*x-k_6*u_x**2*x**2 
     -2*k_7*u_x**2-4*k_8*u_x*x+4*k_9*u*x-2*k_9*u_x*x**2)/4},
  {k_1,k_2,k_3,k_4,k_5,k_6,k_7,k_8,k_9,k_10},
  {}
 }
}
\end{verbatim}
\emph{i.e.}, the expression of the characteristic function $\Omega(x,u,du/dx)$ involves 10 arbitrary constants.
A basis of the 10--dimensional Lie algebra of contact symmetries is obtained by invoking
\begin{verbatim}
reliegen(1,{-1,-1,-1,-1,-1,-1,1,1,1,1}) $
\end{verbatim}
providing
\begin{verbatim}
generators ->
{
 {0, x**2/2, x},
 {0, x, 1},
 {0, u, u_x},
 {0, 1, 0},
 {u - u_x*x, ( - u_x**2*x)/2, ( - u_x**2)/2},
 {(x*(2*u - u_x*x))/2, (4*u**2 - u_x**2*x**2)/4, (u_x*(2*u - u_x*x))/2},
 {u_x, u_x**2/2, 0},
 {x, 0, - u_x},
 {x**2/2, u*x, u},
 {1, 0, 0}
}
\end{verbatim}
Note that only the fifth, sixth and seventh generators correspond to proper contact transformations; the remaining ones are prolongations of point symmetries.
\end{example}

\subsection{Computation of variational symmetries and associated conservation laws}
For computing variational symmetries, besides assigning {\tt jetorder}, {\tt xvar} and {\tt uvar} we need to assign the list 
{\tt lagrangian} with only one element, and set the variable {\tt variational} to 1.
The functions that we have to call do not change, as the following example shows.

\begin{example}
The  code 
\begin{verbatim}
jetorder:=1 $
variational:=1 $
xvar:={t} $
uvar:={u} $
lagrangian:={t^2*u_t^2/2-t^2*u^6/6} $
relie(4) $
\end{verbatim}
allows the user to compute the variational symmetries of the Lagrangian 
\[
\mathcal{L}=\frac{t^2}{2}\left(\frac{du}{dt}\right)^2-\frac{t^2u^6}{6},
\]
corresponding to the Emden--Fowler equation
\[
\frac{d^2u}{dt^2}+\frac{2}{t}\frac{du}{dt}+u^5=0.
\]
We get
\begin{verbatim}
symmetries ->
{
 {{},
 {phi_t=k_2/6, eta_u=(k_1*u)/2, xi_t= - k_1*t},
 {k_1,k_2},
 {}}
}
\end{verbatim}
where, besides the expressions of the infinitesimal generators, we have also the function $\phi$ entering the definition of 
variational symmetries.

Calling {\tt reliegen(1,\{\})}, we obtain the two lists
\begin{verbatim}
generators ->
{
 { - t, u/2}
}
\end{verbatim}
and
\begin{verbatim}
cogenerators ->
{
 {0}
}
\end{verbatim}
\emph{i.e.}, the (list of) Lie generator(s) of the variational symmetries and the (list of) function(s) entering the 
invariance condition of the Lagrangian action.  

We have only a Lie generator, whereas the function $\phi$  can be taken without loss of generality equal to zero.

Finally,  calling {\tt relieclaw(1)}, we get 
\begin{verbatim}
fluxes ->
{
 (t**2*(t*u**6 + 3*t*u_t**2 + 3*u*u_t))/6
}
\end{verbatim}
\emph{i.e.}, the first integral
\[
 \frac{1}{6} \left( t^2 \left( tu^6 + 3t \left(\frac{du}{dt} \right)^2 + 3u\frac{du}{dt}\right)\right)= \hbox{constant}.
\]

\end{example}
\begin{example}[Klein--Gordon equation (see \cite{BlumanCheviakovAnco2009book})]
The class of Klein--Gordon wave equations,
\begin{equation}
\label{KleinGordon} \frac{\partial^2 u}{\partial t\partial x} + g(u) = 0
\end{equation}
with a general nonlinear interaction term $g(u)$, can be derived from a variational
principle given by the action functional with Lagrangian
\begin{equation}
\mathcal{L}=-\frac{1}{2}\frac{\partial u}{\partial t}\frac{\partial u}{\partial x}+h(u), \qquad h^\prime(u)=g(u).
\end{equation}
For an arbitrary $g(u)$, the generators of these symmetries  are
\begin{equation}
{\Xi}_1=t\frac{\partial}{\partial t}-x\frac{\partial}{\partial x}, \qquad {\Xi}_2=\frac{\partial}{\partial t}, \qquad
{\Xi}_3=\frac{\partial}{\partial x}, 
\end{equation}
and it is straightforward to verify that they are variational symmetries of the action functional.
Applying Noether's theorem, the following conservation laws are easily derived:
\begin{equation}
\begin{aligned}
&\frac{D}{D t}\left(\frac{1}{2}\left(\frac{\partial u}{\partial x}\right)^2\right)+\frac{D}{D x}\left(h(u)\right)=0,\\
&\frac{D}{D t}\left(h(u)\right)+\frac{\partial}{\partial x}\left(\frac{1}
{2}\left(\frac{\partial u}{\partial t}\right)^2\right)=0,\\
&\frac{D}{D t}\left(\frac{1}{2}x\left(\frac{\partial u}{\partial x}\right)^2-t
h(u)\right)+\frac{D}{D x}\left(-\frac{1}{2}t\left(\frac{\partial u}{\partial t}\right)^2+x
h(u)\right)=0.
\end{aligned}
\end{equation}
The \reduce code for this computation is
\begin{verbatim}
jetorder:=1 $
xvar:={t,x} $
uvar:={u} $
depend h,u $
lagrangian:={-u_t*u_x/2+h} $
variational:=1 $
relie(4) $
reliegen(1,{-1,-1,-1}) $
\end{verbatim}
The fluxes entering the conservation laws associated to the three Lie generators are obtained with the calls
{\tt relieclaw(1)}, {\tt relieclaw(2)}, and {\tt relieclaw(3)}, respectively.
\end{example}
 
\subsection{Computation of approximate symmetries}
Approximate Lie point symmetries of differential equations containing small terms are computed according to the approach described in \cite{DSGO:lieapprox}. The small parameter must be denoted by {\tt epsilon}. In addition to the input data
described before, the user has to define the integer {\tt approxorder} for the order of approximation, and a rule,
say {\tt let epsilon**(approxorder+1) = 0}. The same functions providing the necessary computations for exact Lie point symmetries work in this case too.
\begin{example}
The following code shows how to compute first order approximate symmetries of the Korteweg--deVries--Burgers equation
\begin{equation}
\label{kdvb}
\frac{\partial u}{\partial t}+u\frac{\partial u}{\partial x}+\frac{\partial^3 u}{\partial x^3}-\varepsilon \frac{\partial^2 u}{\partial x^2}=0
\end{equation}
\begin{verbatim}
jetorder:=3 $
approxorder:=1 $
let epsilon^2 = 0 $
xvar:={t,x} $
uvar:={u} $
diffeqs:={u_t+u*u_x+u_xxx-epsilon*u_xx} $
leadders:={u_t} $
relie(4) $
\end{verbatim}
As a result, we get:
\begin{verbatim}
symmetries ->
{
 {
  {},
  {eta1_u=( - 3*k_4 + 2*k_6*u0)/3,
   eta0_u= - k_1,
   xi1_x=( - 3*k_4*t - 3*k_5 - k_6*x)/3,  
   xi1_t= - k_6*t - k_7,
   xi0_x= - k_1*t - k_2,
   xi0_t= - k_3},
  {k_1,k_2,k_3,k_4,k_5,k_6,k_7},
  {}
 }
}
\end{verbatim}
Invoking {\tt reliegen(1,\{-1,-1,-1,-1,-1,-1,-1\})} provides
\begin{verbatim}
generators ->
{
  {0,t,1},
  {0,1,0},
  {1,0,0},
  {0,epsilon*t,epsilon},
  {0,epsilon,0},
  {epsilon*t,(epsilon*x)/3,( - 2*epsilon*u0)/3},
  {epsilon,0,0}
}
\end{verbatim}
so that a basis of the approximate Lie algebra of symmetries of equation~\eqref{kdvb} is (after a reordering)
\begin{equation}
\begin{aligned}
&\Xi_1=\frac{\partial}{\partial t}, \qquad \Xi_2=\frac{\partial}{\partial x},\qquad \Xi_3=t\frac{\partial}{\partial x}+\frac{\partial}{\partial u},\\
&\Xi_4=\varepsilon \left(t \frac{\partial}{\partial t}+ \frac{x}{3}\frac{\partial}{\partial x}-\frac{2}{3}u_0\frac{\partial}{\partial u}
\right),\qquad
\Xi_5=\varepsilon \left(t\frac{\partial}{\partial x}+\frac{\partial}{\partial u}\right)=\varepsilon\Xi_3,\\
&\Xi_6=\varepsilon\frac{\partial}{\partial t}=\varepsilon\Xi_1, \qquad \Xi_7=\varepsilon \frac{\partial}{\partial x}=\varepsilon \Xi_2.
\end{aligned}
\end{equation}
\end{example}

\begin{remark}
If the differential equation involves some of the dependent variables or some of their derivatives not in explicit and definite form,  the user needs to expand these functions in power series of $\varepsilon$ up to the desired order of approximation. For instance, to compute the first order approximate symmetries of the equation
\[
\frac{\partial^2 u}{\partial t^2}-\frac{\partial}{\partial x}\left(f(u)\frac{\partial u}{\partial x}\right)+\varepsilon \frac{\partial u}{\partial t}=0,
\]
we have to write
\begin{verbatim}
depend f,u0 $
diffeqs:={u_tt-(f+epsilon*df(f,u0)*u1)*u_xx-(df(f,u0)
              +epsilon*df(f,u0,2)*u1)*u_x^2-epsilon*u_t} $
\end{verbatim}
Of course, the user may insert a differential equation completely expanded in power series of $\varepsilon$, but this is not strictly required.  
\end{remark}
To compute approximate conditional, contact or variational symmetries the only difference with respect to the
corresponding \emph{exact} case consists in setting a positive value to {\tt approxorder} and defining the rule
\begin{verbatim}
let epsilon^(approxorder+1) = 0 $
\end{verbatim}
\subsection{Computation of equivalence transformations}
If we consider a class of differential equations, and want to determine equivalence transformations
some more input data are necessary. In particular, in addition to the data already discussed for 
the computation of Lie point symmetries, we need to fix:
\begin{enumerate}
\item the list {\tt arbelem} of the arbitrary elements involved in the differential equations;
\item the integer {\tt arborder} denoting the highest order of the derivatives of the arbitrary elements
with respect to their arguments;
\item the integer {\tt zorder} characterizing the variables the arbitrary elements depend on; for instance, 
if {\tt zorder} is 0, then the arbitrary elements depend at most on the independent and dependent variables;
if {\tt zorder} is 1, then the arbitrary elements depend at most on the independent, dependent variables and
first order derivatives, \ldots.
\end{enumerate}
If we want to remove the dependence of some arbitrary elements on some variables, we need to add 
\emph{auxiliary conditions} to the differential equations at hand.

The infinitesimal generators of the arbitrary elements are automatically computed by the program (so the user is not requested to set them), 
and stored, together with the infinitesimals of independent and dependent variables in the list 
{\tt allinfinitesimals}. The infinitesimal generator of an arbitrary element is denoted by {\tt mu\_}
followed by  the name of the arbitrary element.
Let us illustrate with two different examples how \relie works in such a situation.

\begin{example}\label{equiv2} \cite{OliveriSpeciale2013JMP}
Consider the class $\mathcal{E}(\mathbf{p})$ with $\mathbf{p}= (p_1,p_2,p_3,p_4,p_5,p_6)$ of
systems
\begin{equation}
\label{sis}
\begin{aligned}
&\frac{\partial u_1}{\partial t}+\frac{\partial u_2}{\partial x}+\frac{\partial u_3}{\partial y} =0,\\
&\frac{\partial u_2}{\partial t}+\frac{\partial p_1}{\partial x}+\frac{\partial p_2}{\partial y}-p_5=0,\\
&\frac{\partial u_3}{\partial t}+\frac{\partial p_3}{\partial x}+\frac{\partial p_4}{\partial y}-p_6=0,
\end{aligned}
\end{equation}
where $t$, $x$ and $y$ are the independent variables, $u_1$, $u_2$ and $u_3$ the dependent variables, whereas $p_i\equiv p_i(t,x,y,u_1,u_2,u_3)$ ($i=1,\ldots,6$) stand for arbitrary continuously differentiable functions of the indicated arguments.

\relie determines the equivalence transformations with the code
\begin{verbatim}
jetorder:=1 $
uvar:={u1,u2,u3} $
xvar:={t,x,y} $
arbelem:={p1,p2,p3,p4,p5,p6} $
arborder:=1 $
zorder:=0 $
diffeqs:={u1_t+u2_x+u3_y,
          u2_t+p1_x+p1_u1*u1_x+p1_u2*u2_x+p1_u3*u3_x
              +p2_y+p2_u1*u1_y+p2_u2*u2_y+p2_u3*u3_y-p5,
          u3_t+p3_x+p3_u1*u1_x+p3_u2*u2_x+p3_u3*u3_x
              +p4_y+p4_u1*u1_y+p4_u2*u2_y+p4_u3*u3_y-p6} $
leadders:={u1_t,u2_t,u3_t} $
relie(4) $
\end{verbatim}
As a result we have the infinitesimals:
\begin{verbatim}
symmetries ->
{
 {
  {
   df(f_3,y)+df(f_6,t)+df(f_7,x)
  },
  {
   eta_u3=-df(f_1,t)*u1-df(f_1,x)*u2+df(f_10,t)*u3+df(f_2,x)*u3+f_3-k_1*u3,
   eta_u2=df(f_1,y)*u2+df(f_10,t)*u2-df(f_2,t)*u1-df(f_2,y)*u3+f_7-k_1*u2,
   eta_u1=df(f_1,y)*u1+df(f_2,x)*u1+f_6-k_1*u1,
   xi_y=-f_1,
   xi_x=-f_2,
   xi_t=-f_10,
   mu_p6=-2*df(f_1,t,x)*u2-2*df(f_1,t,y)*u3-df(f_1,t,2)*u1-df(f_1,x,y)*p2 
         -df(f_1,x,y)*p3-df(f_1,x,2)*p1-df(f_1,x)*p5-df(f_1,y,2)*p4
         +df(f_10,t,2)*u3+2*df(f_10,t)*p6+df(f_2,x)*p6+df(f_3,t)
         +df(f_4,x)-df(f_5,y)-k_1*p6,
   mu_p5=df(f_1,y)*p5+df(f_10,t,2)*u2+2*df(f_10,t)*p5-2*df(f_2,t,x)*u2
         -2*df(f_2,t,y)*u3-df(f_2,t,2)*u1-df(f_2,x,y)*p2
         -df(f_2,x,y)*p3-df(f_2,x,2)*p1-df(f_2,y,2)*p4-df(f_2,y)*p6
         +df(f_7,t)+df(f_8,y)+df(f_9,x)-k_1*p5,
   mu_p4=-2*df(f_1,t)*u3-df(f_1,x)*p2-df(f_1,x)*p3-df(f_1,y)*p4
         +2*df(f_10,t)*p4+df(f_2,x)*p4-f_5-k_1*p4,
   mu_p3=-df(f_1,t)*u2-df(f_1,x)*p1+2*df(f_10,t)*p3-df(f_2,t)*u3
         -df(f_2,y)*p4+f_4-k_1*p3,
   mu_p2=-df(f_1,t)*u2-df(f_1,x)*p1+2*df(f_10,t)*p2-df(f_2,t)*u3
         -df(f_2,y)*p4+f_8-k_1*p2,
   mu_p1=df(f_1,y)*p1+2*df(f_10,t)*p1-2*df(f_2,t)*u2-df(f_2,x)*p1
         -df(f_2,y)*p2-df(f_2,y)*p3+f_9-k_1*p1
  },
  {
    f_1, f_2, f_3, f_4, f_5, f_6, f_7, f_8, f_9, f_10, k_1
  },
  {}
 }
}
\end{verbatim}
where {\tt f\_1}, \ldots, {\tt f\_10} are functions of $t,x,y$, along with the constraint
\begin{equation}
\frac{\partial f_6}{\partial t}+\frac{\partial f_3}{\partial x}+\frac{\partial f_7}{\partial y}=0.
\end{equation}
\end{example}

\begin{example}
The equivalence transformations of the equation
\begin{equation}
\frac{\partial^2 u}{\partial t^2}-f\left(x,\frac{\partial u}{\partial x}\right)\frac{\partial^2 u}{\partial x^2}-
g\left(x,\frac{\partial u}{\partial x}\right)=0
\end{equation}
are immediately found in \relie with the following code:
\begin{verbatim}
jetorder:=2 $
xvar:={t,x} $
uvar:={u} $
zorder:=1 $
arborder:=1 $
arbelem:={f,g} $
diffeqs:={u_tt-f*u_xx-g, f_t, g_t, f_u, g_u, f_u_t, g_u_t} $
leadders:={u_tt, f_t, g_t, f_u, g_u, f_u_t, g_u_t} $
relie(4) $
\end{verbatim}
Notice that we have added in {\tt diffeqs} the auxiliary conditions; the variable {\tt zorder} has been set to 1 since the functions $f$ and $g$ depend on first order derivatives; moreover, the variable {\tt arborder} is set to 1 since in the auxiliary conditions the derivatives of $f$ and $g$ with respect to some of their arguments occur. 

As a result we obtain:
\begin{verbatim}
symmetries ->
{
 {
  {},
  {eta_u=(2*f_1-2*k_1*u-k_6*t**2-2*k_7*t-2*k_8)/2,
   xi_x= -k_4*x-k_5,
   xi_t=-k_2*t-k_3,
   mu_g=-df(f_1,x,2)*f-g*k_1+2*g*k_2-k_6,
   mu_f=2*f*k_2-2*f*k_4
  },
  {f_1,k_1,k_2,k_3,k_4,k_5,k_6,k_7,k_8},
  {}
 }
}
\end{verbatim}
where the function {\tt f\_1} is arbitrary and depends on $x$.
\end{example}

\begin{remark}
By default, the infinitesimals for the independent and dependent variables do not depend on arbitrary elements; if we are interested to 
general equivalence transformations where also the infinitesimal generators of the independent and dependent variables depend on the 
arbitrary elements \cite{Meleshko:generalequivalence}, then we have to add the
statement
\begin{verbatim}
generalequiv:=1 $
\end{verbatim}
\end{remark}

\subsection{Inverse Lie problem}
Here we show with a simple example how \relie can be used for investigating the Lie remarkability of differential equations \cite{MOV2007,MOV2007b,MOSV2014,GO2019_Lierem}. The following code shows how one can easily verify that Monge--Amp\`ere equation describing a surface in $\mathbb{R}^3$ with zero Gaussian curvature,
\[
\frac{\partial^2 u}{\partial x^2}\frac{\partial^2 u}{\partial y^2}-\left(\frac{\partial^2 u}{\partial x\partial y}\right)^2=0,
\] 
is strongly Lie remarkable \cite{MOV2007}:
\begin{verbatim}
jetorder:=2 $
xvar:={x,y} $
uvar:={u} $
relieinit() $
generatealgebra(2) $
\end{verbatim}
After setting the necessary objects ({\tt jetorder}, {\tt xvar} and {\tt uvar}), and calling {\tt relieinit()}, we generate the vector fields spanning the affine Lie algebra in $\mathbb{R}^3$.
Then, calling {\tt testrank(generators)}, we ascertain that the rank of the second order distribution is maximal (equal to the dimension of second order jet space); the call {\tt inverselie(8)}  computes the list {\tt allminors} of all minors of order 8 of such a distribution; all minors vanish by evaluating them on the differential equation, that can be verified by calling
\begin{verbatim}
allminors:=sub(u_yy=u_xy^2/u_xx,allminors) $
\end{verbatim}

As a last check, 
\begin{verbatim}
rank(sub(u_yy=u_xy^2/u_xx,distribution);
\end{verbatim}
returns 7, which is the dimension of the manifold in the second order jet space characterized by Monge--Amp\`ere equation.

\section{Inside \relie: global variables and routines}
\label{sec:insiderelie}
\relie uses some global variables to perform the various tasks: they can be distinguished among \emph{input variables}
(that the user needs to set before starting computation), \emph{output variables} (computed by \relie and of interest to the user), and \emph{intermediate variables} (computed by \relie and in general not of interest to the common user).

\subsection{Input variables} 
\begin{itemize}

\item {\tt approxorder}: maximum order of approximate symmetries of equations containing a small parameter; by default it is $0$, \emph{i.e.}, exact symmetries; the small parameter involved in the approximate symmetries must be denoted by {\tt epsilon}; when looking for approximate symmetries the user has to define the rule {\tt let epsilon**(approxorder+1)=0};

\item {\tt arbelem}: list of the arbitrary elements (only for equivalence transformations); by default it is an 
empty list;

\item {\tt arborder}: maximum order of derivatives of arbitrary elements (only for equivalence 
transformations); by default it is $-1$, \emph{i.e.}, point symmetries; 

\item{\tt contact}: set to 1 for contact symmetries; by default it is 0;

\item {\tt diffeqs}: list of the left--hand sides of differential equations (with vanishing right--hand sides);

\item {\tt freepars}: list of arbitrary constants or functions involved in the differential equations (for group classification problems);  by default it is an empty list;

\item {\tt generalequiv}:  set to 1 for general equivalence transformations where all infinitesimals depend 
on independent and dependent variables and arbitrary elements; the default value is 0, meaning that
the infinitesimals of independent and dependent variables do not depend upon the arbitrary elements;

\item {\tt jetorder}:  maximum order of derivatives in differential equations;

\item {\tt lagrangian}: a list with only one element corresponding to the Lagrangian (it is necessary to set 
{\tt variational} to 1);

\item {\tt leadders}: list of the leading derivatives; {\tt diffeqs} are solved with respect to them;

\item {\tt nonclassical}: set to a value between 1 and the number of independent variables (only for 
conditional symmetries); by default it is 0, \emph{i.e.}, classical symmetries;

\item {\tt nonpolyders}: list of derivatives not occurring in polynomial form in the differential equations; by default it is an empty list;

\item {\tt nonzeropars}: list of arbitrary constants or functions involved in the differential equations that can not vanish;  by default it is an empty list;

\item {\tt qcond}: list of indexes of dependent variables whose invariant surface conditions have to be used 
for computing conditional symmetries;

\item {\tt uvar}: list of the dependent variables;

\item{\tt variational}: set to 1 if variational symmetries of a Lagrangian are needed; by default it is 0;

\item {\tt xvar}: list of the independent variables;

\item {\tt zorder}: maximum order of derivatives of {\tt uvar} with respect to {\tt xvar} the elements in {\tt 
arbelem} depend on (only for equivalence transformations); if {\tt zorder} is set to 0, the arbitrary elements 
depend on {\tt xvar} and {\tt uvar}; {\tt zorder} cannot exceed {\tt jetorder}.

\end{itemize}

\subsection{Output variables} 
\begin{itemize}

\item {\tt allinfinitesimals}: list of two lists; the first sublist is the list of the infinitesimals, in order, of the 
independent variables, dependent variables and arbitrary elements (the latter in the case of equivalence 
transformations); the second sublist is the list of various terms (constants and functions) involved in the 
expression of infinitesimals;

\item {\tt allminors}: list of minors of a given order extracted from a matrix; returned by the function {\tt minors(m,k)}, where {\tt m} is a matrix and {\tt k} a positive integer, or by the function {\tt inverselie(k)} that takes the {\tt jetorder}-th distribution of {\tt generators} as the matrix from which the minors of order $k$ are extracted;
  
\item {\tt arbconst}: list of arbitrary constants involved in the expression of the symmetries;

\item {\tt arbfun}: list of arbitrary functions involved in the symmetries;

\item {\tt cogenerators}: list of the functions entering the definition of variational symmetries
and corresponding to the infinitesimal generators (produced by {\tt reliegen()});

\item {\tt commtable}: table of commutators of a list of vector fields;

\item {\tt deteqs}: list of the determining equations (produced by {\tt reliedet()});

\item {\tt distribution}: matrix of the {\tt jetorder}-th distribution of a list of generators (produced by {\tt reliedistrib()}), 
\emph{i.e.}, a matrix where each row is the prolonged vector field evaluated on one of the provided 
infinitesimal generators;

\item {\tt fluxes}: list of the  fluxes of the conservation law corresponding to a Lie generator (computed by
{\tt relieclaw()});

\item {\tt generators}: list of the infinitesimal generators of the finite Lie algebra admitted by the differential 
equations at hand (produced by {\tt reliegen()}); the list  {\tt generators}
may also been obtained by calling {\tt generatealgebra(k)}, where {\tt k} can be 1 (algebra of isometries), 2 (algebra of affine transformations) or 3 (algebra of projective transformations); of course, it is necessary to set 
{\tt jetorder}, {\tt xvar} and {\tt uvar} before calling {\tt generatealgebra(k)};
 
\item {\tt invcond}: list of the invariance conditions of the differential equations at hand (produced by {\tt 
relieinv()});

\item {\tt nzcomm}: list  of non--zero commutators of a list of vector fields;

\item {\tt prolongation}: list of two lists: the first one is the list of the coordinates of the jet space, the second 
one the list of the corresponding infinitesimals (produced by {\tt relieprol()});

\item {\tt splitsymmetries}: list of lists: the $k$-th element is a list containing the infinitesimals corresponding to the $k$-th element of {\tt generators} (produced by {\tt reliegen()}); the list  {\tt splitsymmetries}
may also been obtained by calling {\tt generatealgebra(k)}, where {\tt k} can be 1 (algebra of isometries), 2 (algebra of affine transformations) or 3 (algebra of projective transformations); of course, it is necessary to set 
{\tt jetorder}, {\tt xvar} and {\tt uvar} before calling {\tt generatealgebra(k)}; the list {\tt splitsymmetries} is used internally by the functions {\tt reliedistrib()}, {\tt inverselie()} and {\tt testrank()};

\item {\tt symmetries}: list of the infinitesimals admitted by the differential equations at hand (produced 
by {\tt reliesolve()}).

\end{itemize}

\subsection{Intermediate variables}
\begin{itemize}

\item {\tt jet}: 
list of three lists: indices for computing the infinitesimals and their prolongations, coordinates of jet space 
and their internal representation;

\item {\tt jetapprox}: 
list of two lists: list of independent variables and expansions of dependent variables and their derivatives, 
and list of their internal representation (only for approximate symmetries);

\item {\tt jetequiv}: 
list of three lists: indices for computing the infinitesimals and their prolongations for arbitrary elements, 
arbitrary elements, and their internal representation (only for equivalence transformations);

\item {\tt jetsplit}:
list of two lists: indices for computing the infinitesimals and their prolongations, list of independent variables, 
zeroth order dependent variables and their derivatives (only for approximate symmetries);

\item {\tt solutiondedv}:
solution of the differential equations specified in {\tt diffeqs} with respect to {\tt leadders}; for conditional 
symmetries, the invariant surface conditions and their needed differential consequences are solved too;

\item {\tt steprelie}: integer that stores the status of the computation; 0: no computation done; 
1: {\tt relieinit()} has been called; 2: {\tt relieinv()} has been called; 3: {\tt reliedet()} has been called; 
4: {\tt reliesolve()} has been called;  

\item {\tt zvar}: 
list of two lists: the first one is the list of the variables {\tt arbelem} depend on, the second one the 
corresponding infinitesimals (only for equivalence transformations). 

\end{itemize}

\subsection{Functions}
A short description of  the main functions the user may call is as follows:
\begin{itemize}

\item {\tt abelian(gens)}: checks if the generators {\tt gens} span an Abelian Lie algebra;

\item {\tt commutatortable(gens)}: returns the commutator table of the generators {\tt gens};

\item {\tt essentialpars(gens,vars)}: takes a list of generators {\tt gens} of a multiparameter Lie group of 
transformations for the variables {\tt vars}, and returns the generators which are not linearly independent;

\item {\tt generatealgebra(k)}: once {\tt jetorder}, {\tt xvar} and {\tt uvar} have been properly assigned, this function returns a list of generators spanning the algebra of isometries (for {\tt k = 1}), affine algebra (for {\tt k = 2}), projective algebra (for {\tt k = 3});

\item {\tt inverselie(k)}: computes all the minors of order {\tt k} of the {\tt jetorder}-th distribution generated by the list of vector fields contained in {\tt generators};
  
\item {\tt liebracket(gen1,gen2)}: returns the Lie bracket of the generators {\tt gen1} and {\tt gen2};

\item {\tt newordering(lis,ind)}: returns a list of the elements in {\tt lis} reordered according to the permutation
{\tt ind} of the integers $\{1,\ldots,n\}$, where $n$ is the length of list {\tt lis};

\item {\tt nonzerocommutators(gens)}:  returns {\tt nzcomm}, a list of non--zero 
commutators of generators {\tt gens};

\item {\tt offprintcrack()}: prevents {\tt reliesolve()} to display the steps needed for solving determining equations (this is the default configuration);

\item {\tt onprintcrack()}: sets a variable used in Crack package (used in the function {\tt reliesolve()}) in order to display the steps needed for solving determining equations;

\item {\tt relieclaw(k)}:  returns {\tt fluxes}, a list of the components of the fluxes of the conservation law
corresponding to the {\tt k}-th Lie generator (obtained after calling to {\tt reliegen()});

\item {\tt reliedet()}: splits the invariant conditions providing the list {\tt deteqs} of determining equations;

\item {\tt reliedistrib()}: returns the matrix {\tt distribution}, \emph{i.e.}, 
a matrix whose rows are the prolonged vector fields evaluated in the list {\tt splitsymmetries} (computed by the function
{\tt reliegen()}, or by the function {\tt generatealgebra()}, or suitably assigned by the user; 

\item {\tt reliegen(k,lis)}: returns the list {\tt generators}; $k$ is an integer (less or equal to the length of {\tt 
symmetries}) and {\tt lis} a list that 
can be empty;  if {\tt lis} is made by as many values as the number of arbitrary constants occurring in {\tt symmetries}, {\tt generators}
consists of a list of vector fields, where each vector field is obtained replacing the $i$-th parameter by the $i$-th element in the list {\tt lis} 
(or 1 if {\tt lis} is empty or its length is different from the number of arbitrary constants entering {\tt 
symmetries}) and the other parameters are replaced by 0; if the list {\tt lis} is {\tt \{-1\}}, then
{\tt generators} is a list with only one element containing the components of the infinitesimals in their general form, \emph{i.e.}, the linear combinations of all admitted generators; the function produces also the list {\tt splitsymmetries} whose $k$-th element is a list containing the infinitesimals corresponding to the $k$-th element of {\tt generators};

\item {\tt relieinit()}: if input data have been correctly defined, the function initializes the objects
for doing the computation;

\item {\tt relieinv()}: computes the invariance conditions; returns {\tt invcond};

\item {\tt relieprol()}: returns the prolongation of a general vector field;

\item {\tt reliesolve()}: solves the determining equations for the infinitesimals; 
returns {\tt symmetries}; in group classification problems 
(but also in the case of conditional symmetries), the list {\tt symmetries} may contain different solutions for 
the infinitesimals according to the values of {\tt freepars}; as a default {\tt reliesolve()} does not display the steps made to obtain the solution of determining equations; the user can see these steps by calling {\tt onprintcrack()}; this is suggested when {\tt reliesolve()} seems to use too much time to complete its execution; the list {\tt symmetries} contains a list of the sets of solutions of determining equations; each element of this list in turn is a list of four elements: the first one is a list of conditions (possibly empty) that remained unsolved; the second one is a list giving the solution to the determining equations, \emph{i.e.}, the expressions of the infinitesimals;
the third one is a list containing the parameters involved in the solution; the fourth one is a list of expressions which can not vanish (this list can be empty);

\item {\tt solvable(gens)}: checks if the generators {\tt gens} span a solvable Lie algebra;

\item {\tt testrank(gens)}: returns the rank of the {\tt jetorder}-th distribution generated by the generators {\tt gens}.

\end{itemize}

Notice that \relie contains some companion functions, used by the main procedures. Here we list some of 
them that can be useful in interactive sessions.
\begin{itemize}

\item {\tt allcoeffs(lis1,lis2)}: returns the list of coefficients of {\tt lis1} (list of polynomyals) 
with respect to the variables in list {\tt lis2}; 

\item {\tt bincoeff(n,k)}: returns $\binom{n}{k}$;

\item {\tt combnorep(n,k)}: returns the combinations without repetition of $k$ elements chosen in $\{1,2,\ldots,n\}$;

\item {\tt combrep(n,k)}: returns the combinations with repetition of $k$ elements chosen in $\{1,2,\ldots,n\}$;

\item {\tt delzero(lis)}: returns a list containing all non--zero elements in the list {\tt lis};

\item {\tt dependence(lis1,lis2)}: declares that the elements in the list {\tt lis1} depend on the variables in the 
list {\tt lis2};

\item {\tt dlie(obj,var)}: returns the usual Lie derivative of {\tt obj} with respect to {\tt var};

\item {\tt dlieapprox(obj,var)}: returns the Lie derivative of {\tt obj} with respect to {\tt var} in the context of approximate symmetries;

\item {\tt dliestar(obj,var)}: returns the additional Lie derivative used for equivalence transformations;

\item {\tt kroneckerdelta(k1,k2)}: returns 1 if $k1=k2$, 0 otherwise;

\item {\tt letterlist(obj,n)}: {\tt obj} is a symbol, $n$ a positive integer;  \\
for instance, {\tt letterlist(x,4)} builds the list  $\{x1,x2,x3,x4\}$;

\item {\tt letterlistvar(obj,lis)}: {\tt obj} is a symbol, {\tt lis} a list; \\
for instance, {\tt letterlistvar(xi\_,\{x,y\})} builds the list {\tt \{xi\_x,xi\_y\}};

\item {\tt listletter(lis,ch)}: {\tt lis} is a list, {\tt ch} a symbol; \\
for instance, {\tt listletter(\{u\_,v\_\},x)} builds the list {\tt \{u\_x,v\_x\}};

\item {\tt membership(elem,lis)}: returns the number of occurrences of the element {\tt elem} in the list {\tt lis};

\item {\tt minors(m,k)}: {\tt m} is a matrix and {\tt k} is a positive integer: returns the list of minors of
order {\tt k} of the matrix {\tt m};

\item {\tt nodependence(lis1,lis2)}: removes the dependence of the objects in the list {\tt lis1} upon the 
variables in the list {\tt lis2};

\item {\tt removeelement(lis,elem)}: removes the element {\tt elem} from the list {\tt lis};

\item {\tt removemultiple(lis1,lis2)}: removes from the list {\tt lis1} the elements of the list {\tt lis2};

\item {\tt scalarmult(obj,lis)}: returns a list whose $k$-th element is the product of the scalar {\tt obj} and the 
$k$-th element of list {\tt lis};

\item {\tt scalarproduct(lis1,lis2)}: returns the sum of the products element by element of
two lists with the same number of elements;

\item {\tt sumlist(lis1,lis2)}: returns a list summing element by element the two lists with the same length;

\item {\tt zerolist(n)}: returns a list of $n$ zeros, the empty list if $n\le 0$.

\end{itemize}

\section{Conclusions}
The program \relie allows the user to perform almost automatically much of the computations needed for investigating ordinary and partial differential equations by means of Lie group methods. The package can also be used in interactive
computations where some special assumptions need to be made. Remarkably, the program works in a CAS like 
\reduce which is open source and freely available for all operating systems. The various routines of \relie have been
tested along the years in many different situations and hopefully are sufficiently reliable. 
At the url \url{http://mat521.unime.it/oliveri} the source code of the package, together with the user's manual, can be found. 

We illustrated the use of the package by choosing some rather simple examples in different areas of Lie symmetry analysis of differential equations; nevertheless, this does not mean that only simple problems can be faced. In fact, \relie provides effective also when one has to investigate systems of partial differential equations that require to deal with very long expressions. As a last remark, we observe that
this is the first program able to compute approximate Lie symmetries of differential equations containing small terms according to the theory recently proposed \cite{DSGO:lieapprox},
as well as approximate Q-conditional symmetries 
\cite{GorgoneOliveriEJDE2018,GO-ZAMP}.

\section*{Acknowledgements}
Work supported by G.N.F.M. of ``Istituto Nazionale di Alta Matematica''. 
The author is grateful to colleagues and friends that used along the years part of this 
program in their research activity so contributing to fix bugs. 
In particular, the author warmly thanks dr. Matteo Gorgone for many useful suggestions and criticisms.


\end{document}